\documentstyle[preprint,aps,eqsecnum,prabib]{revtex}
\tighten
\begin{document}

\input{BoxedEPS.tex}

\SetRokickiEPSFSpecial
\HideDisplacementBoxes
%\SetEPSFDirectory{Figures/}

\title{Electron-deuteron scattering in a current-conserving
description of relativistic bound states: formalism and impulse approximation 
calculations}

\author {D.~R. Phillips and S.~J. Wallace
\footnote{Email: phillips@quark.umd.edu, wallace@quark.umd.edu.}}

\address{Department of Physics and Center for Theoretical Physics,\\
University of Maryland, College Park, MD, 20742-4111}

\author{N.K. Devine\footnote{Email: devine@quark.umd.edu}}

\address{\it General Sciences Corporation, Bowie, MD}

\date{\today}
\maketitle

\begin{abstract}
The electromagnetic interactions of a relativistic two-body bound
state are formulated in three dimensions using an equal-time (ET)
formalism. This involves a systematic reduction of four-dimensional
dynamics to a three-dimensional form by integrating out the time
components of relative momenta. A conserved electromagnetic current is
developed for the ET formalism. It is shown that consistent
truncations of the electromagnetic current and the $NN$ interaction
kernel may be made, order-by-order in the coupling constants, such
that appropriate Ward-Takahashi identities are satisfied.  A
meson-exchange model of the $NN$ interaction is used to calculate
deuteron vertex functions. Calculations of electromagnetic form
factors for elastic scattering of electrons by deuterium are performed
using an impulse-approximation current. Negative-energy components of
the deuteron's vertex function and retardation effects in the
meson-exchange interaction are found to have only minor effects on the
deuteron form factors.

\end{abstract}

\section{Introduction and Summary of results}
\label{intro-summary}

Experiments being performed at the Thomas Jefferson National
Accelerator Facility (TJNAF) are designed to test our understanding of
deuteron properties at space-like momentum transfers comparable to the
nucleon mass. In building theoretical models of these processes,
relativistic kinematics and dynamics would seem to be called for.
Hence, considerable theoretical effort has been invested in
constructing relativistic formalisms for the two-nucleon bound state
that are based on an effective quantum-field-theory Lagrangian. If
the usual hadronic degrees of freedom appear in the Lagrangian, then
this strategy can be used to obtain a logical extension of the
standard non-relativistic treatment of the two-nucleon system.  The
central goal of such an approach is to capture the relativistic
effects in the two-nucleon system. At the same time, for applications
to electromagnetic physics, it is critical to embed the Ward-Takahashi identities
that imply current conservation for electromagnetic interactions in
the theory.

This paper focuses on a three-dimensional (3D) formalism that, in
principle, is equivalent to the four-dimensional (4D) Bethe-Salpeter
formalism. This approach has been developed in two recent
papers~\cite{PW96,PW97}. We review the formalism for relativistic
bound states, and provide an extension that ensures nonsingular
behavior of the interaction in frames where the total momentum of the
bound system is nonzero.  We also construct the corresponding
electromagnetic current.  Calculations of elastic electron-deuteron
scattering are performed based upon the impulse approximation and the
results for the observables $A$, $B$ and $T_{20}$ are presented.

The Bethe-Salpeter equation, 
\begin{equation}
T=K + K G_0 T,
\label{eq:BSE}
\end{equation}
for the four-dimensional $NN$ amplitude $T$ provides a theoretical
description of the deuteron that includes relativity. In
Eq.~(\ref{eq:BSE}), $K$ is the Bethe-Salpeter kernel, and $G_0$ is the
free two-nucleon propagator.  In a strict quantum-field-theory
treatment, the kernel $K$ includes the infinite set of two-particle
irreducible $NN \rightarrow NN$ Feynman graphs.  Numerical methods to
calculate the 4D t-matrix of 
Eq.~(\ref{eq:BSE}), with an infinite kernel containing all crossed
ladders, have been developed and demonstrated for a simple scalar
field theory by Nieuwenhuis and Tjon~\cite{NT96}.

For the two-nucleon system, such an application of the full
effective quantum field theory of nucleons and mesons is not only
impractical, it also is inappropriate because hadronic degrees
of freedom are not fundamental.  Rather, the Bethe-Salpeter formalism
serves as a theoretical framework within which a relativistic
effective interaction may be developed. This is entirely analogous to
the way that the Schr\"{o}dinger equation serves as a framework for
development of a non-relativistic potential that describes the $NN$
phase shifts.  Since the $NN$ interaction is an
effective one, it is equally appropriate to develop the relativistic
effective interaction within an equivalent three-dimensional formalism
that is obtained from the four-dimensional Bethe-Salpeter formalism
via a systematic reduction technique.

In Ref.\cite{PW96}, a 3D reduction of the Bethe-Salpeter formalism
was developed such that the resulting equations involve the same
propagator as appears in the Salpeter equation~\cite{Sa52},
\begin{equation}
T_1=K_{1} + K_{1} \langle G_0 \rangle  T_1.
\label{eq:Salpeter}
\end{equation}
The three-dimensional propagator $\langle G_0 \rangle$ is obtained by
integrating over the time-component of relative momentum,
\begin{equation}
\langle G_0 \rangle=\int \frac{dp_0}{2 \pi} G_0(p;P).
\label{eq:<G0>}
\end{equation}
The 3D kernel is defined by solving two coupled equations,
\begin{equation}
K_1 = \langle G_0 \rangle ^{-1} \langle G_0 K {\cal G} \rangle
\langle G_0 \rangle ^{-1} ,
\label{eq:K1}
\end{equation}
which is three-dimensional, and 
\begin{equation}
{\cal G} = G_0 + G_0 (K - K_1) {\cal G},
\label{eq:calG-G0}
\end{equation}
which is four dimensional. In fact, the condition (\ref{eq:K1}) is
obtained by demanding that $\langle {\cal G} \rangle = \langle G_0
\rangle$. (See Ref.~\cite{PW96} for details.
Other works which have considered a similar formalism include
Refs.~\cite{Kl53,Kl54,LT63,KL74,LA97}.)  This formalism is systematic
in the sense that, given a perturbative expansion for the 4D kernel,
$K$, a perturbative expansion for the 3D kernel, $K_1$, can be
developed.  However, it is necessary for $\langle G_0 \rangle$ to be
invertible in order for the 3D reduction to contain the full content
of the 4D theory.

\begin{figure}[h]
\centerline{\BoxedEPSF{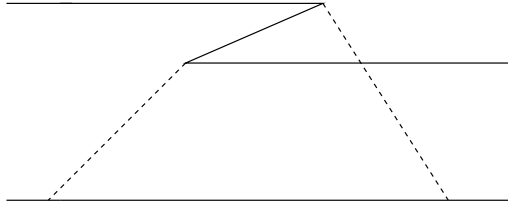 scaled 500}}
\vskip 5 mm
\caption{One example of a Z-graph which is not included in the ladder
Bethe-Salpeter equation scattering series}
\label{fig-Zgraph}  
\end{figure}

As mentioned above, when $G_0$ is the standard two-particle propagator
of the Bethe-Salpeter formalism, $\langle G_0 \rangle$ is the 3D
Salpeter propagator.  For spin-1/2 particles it is,
\begin{equation}
\langle G_0 \rangle=\frac{\Lambda_1^+ \Lambda_2^+}{P^0 - \epsilon_1
- \epsilon_2} - \frac{\Lambda_1^- \Lambda_2^-}{P^0 +  \epsilon_1
+ \epsilon_2};
\label{eq:aveG0}
\end{equation}
where $\Lambda^{\pm}$ are related to projection operators onto
positive and negative-energy states of the Dirac equation, $P^0$ is the
total energy of the two-body system, and
\begin{equation}
\epsilon_i=({\bf p}_i^2 + m_i^2)^{1/2}.
\end{equation} 
Here, and throughout the paper, these single-particle energies are to
be understood as having an infinitesimal negative imaginary part. This
defines our $i \epsilon$ prescriptions. Note that for spin-half
particles, the equal-time propagator of the Salpeter equation is
defined only on two of the four sectors of the Dirac space of two
particles.  It is not invertible.  Although consistent equations can
be projected out for the $++$ and $--$ components, one must set the
$+-$ and $-+$ components to zero for consistency.  Consequently, the
3D reduction does not have the full content of the 4D theory. If the
4D and 3D theories are to have the same dynamical content then we must
include graphs involving $+-$ states, such as the the time-ordered
Z-graph of Fig.~\ref{fig-Zgraph}, in both. This graph is not contained
in the ladder Bethe-Salpeter equation scattering series.

The absence of this mechanism from the ladder Bethe-Salpeter equation
is related to the fact that if we allow one particle's mass to tend to
infinity, the propagator does not reduce to the Dirac propagator for
the other particle. Consequently, Eq.~(\ref{eq:Salpeter}) does not
possess the correct one-body limit if any finite set of graphs is
chosen for $K_1$.

If a scattering equation with a kernel which contains only a finite
number of graphs is to possess the correct one-body limit, two
distinct criteria must be satisfied.  First the 3D propagator should
limit to the one-body propagator for one particle (either Dirac or
Klein-Gordon, depending upon the spin) as the other particle's mass
tends to infinity. Second, as either particle's mass tends to
infinity, the equation should become equivalent to one in which the
interaction, $K_1$, is static.

In fact, Eq.~(\ref{eq:Salpeter})'s lack of either of these properties
springs directly from the Bethe-Salpeter equation (\ref{eq:BSE}) not
having the correct one-body limit if the kernel does not include
the infinite set of crossed-ladder graphs. As alluded to
above, solution of the Bethe-Salpeter equation with such a kernel is
impractical in the $NN$ system. Nevertheless, the desired properties 
%of crossed-ladder graphs to the kernel may be included in an integral
may be obtained by using a 4D integral equation for $K$ to reorganize
the contributions to the kernel of the Bethe-Salpeter equation
as follows, 
\begin{equation}
K = U + U G_C K. 
\label{eq:U-4D} 
\end{equation}
Given a suitable choice for $G_C$, this equation defines a reduced
kernel, $U$, in terms of the original kernel, $K$.  The propagator, $G_C$,
is chosen so as to separate the parts of the kernel, $K$, that are
necessary to obtain the one-body limits from the parts that are not.
Once this is done, $U$ may be truncated at any desired order without
losing the one-body limits. It is readily seen that the original kernel
$K$ may be eliminated in favor of the reduced one, $U$, so as to obtain the
following 4D equation for the t-matrix, which is equivalent to 
Eqs.~(\ref{eq:BSE}) and (\ref{eq:U-4D}),
\begin{equation}
T = U + U {\cal G}_0 T,
\label{eq:T(U)4D}
\end{equation}  
where 
\begin{equation}
{\cal G}_0 \equiv G_0 + G_C.
\label{eq:calG0}
\end{equation}
We are now in a position to remedy the defects of our previous
three-dimensional reduction. Applying the same 3D reduction procedure
as above to this new 4D equation produces
\begin{equation}
T_1 = U_1 + U_1 \langle {\cal G}_0 \rangle T_1, 
\label{eq:3DET}
\end{equation}
where the 3D propagator is 
\begin{equation}
\langle {\cal G}_0(P) \rangle=\frac{ \Lambda_1^+ \Lambda_2^+}
{{P^0} - \epsilon_1 - \epsilon_2} 
- \frac{ \Lambda_1^+ \Lambda_2^-}
{2 \kappa_2^0 - {P^0} + \epsilon_1 + \epsilon_2} 
- \frac{ \Lambda_1^- \Lambda_2^+}
{{P^0} - 2 \kappa_2^0 + \epsilon_1 + \epsilon_2} 
- \frac{ \Lambda_1^- \Lambda_2^-}{{P^0} + \epsilon_1 + \epsilon_2},
\label{eq:aveG0GC}
\end{equation}
and $\kappa_2^0$ is a parameter that arises from the eikonal
approximation. This 3D propagator is found by integrating the
four-dimensional propagator over the time component of relative
momentum. In configuration space this action is equivalent to
considering the propagator in which the two particles involved are
considered only on equal time slices. Thus this will be referred to as
the ET propagator. It was derived for use with instant interactions by
Mandelzweig and Wallace~\cite{MW87,WM89} with the choice $\kappa_2^0 =
P^0/2 - (m_1^2 - m_2^2)/(2P^0)$.  It has the correct one-body limits as
either particle's mass tends to infinity and has an invertible form.
Meanwhile, the 3D kernel $U_1$ is systematically defined by two
equations that are obtained from Eqs.~(\ref{eq:K1}) and
(\ref{eq:calG-G0}) by the substitutions $K \rightarrow U$, $K_1
\rightarrow U_1$ and $G_0
\rightarrow {\cal G}_0$.   

Equation (\ref{eq:3DET}) is a 3D scattering equation that incorporates
relativistic effects systematically and has the correct one-body
limit. Numerical calculations by Nieuwenhuis and Tjon~\cite{NT96}, and
in our previous paper~\cite{PW96}, suggest that the three-dimensional
equation with a lowest-order kernel provides a good approximation to the relevant physics of the
full scattering series of ladders and crossed ladders.  

Negative-energy states and thus Z-graphs, such as the one shown in
Fig.~\ref{fig-Zgraph}, are included in the 3D formalism in a way that
is symmetrical with respect to interchange of particle labels.  When
the terms in the propagator involving $\Lambda _2 ^-$ are omitted, it
takes a form that is very similar to the 3D propagator of the
spectator formalism of Gross~\cite{Gr69}, as derived with particle 1 on mass
shell.  Correspondingly, when the terms in the propagator involving
$\Lambda _1 ^-$ are omitted, the propagator is very similar to that of
the spectator formalism, derived with particle 2 on mass
shell. Indeed, in $++$ sectors, Eq.~(\ref{eq:aveG0GC}) is the same as
either of these spectator propagators, and thus differences arise only
because of the negative-energy states.  Usually for the $NN$ system, a
symmetrized form of the spectator propagator is used, and this is
obtained by averaging the propagators derived with particle 1 and
particle 2 on mass shell~\cite{Gr92}. However, because of averaging, the
symmetrized spectator propagator has half the propagation amplitudes
of the ET propagator in $+-$ and $-+$ states.  A similar
comparison may be made with the BSLT quasipotential propagator 
of Blankenbecler-Sugar~\cite{BbS66} and
Logunov-Tavkhelidze~\cite{LT63}
that has been used in the work of Hummel
and Tjon~\cite{HT89,HT90,HT94}.  The BSLT propagator also has half
the propagation amplitudes of the ET propagator in $+-$ and $-+$ states.
Because similar couplings to the $+-$ and $-+$ states are
present in all of these approaches when similar meson exchange 
interactions are used, we would expect
the role of negative-energy states to be larger 
when the ET propagator is used that when either the spectator propagator
of Gross or the BSLT propagator of Hummel and Tjon is used.  

A straightforward analysis of time-ordered perturbation theory graphs
for the $NN$ t-matrix in the static limit (see
Appendix~\ref{ap-ZgraphTOPT}) shows that the leading Z-graphs
involving the intermediate $+-$ and $-+$ states are correctly given by
the ET propagator.  Results based upon either the symmetrized Gross
propagator or the BSLT propagator are too small by a factor of two.

 Our preference for the ET formalism is based on three facts:
it embeds the correct one-body limit for {\it either} particle as 
the mass of the other particle tends to infinity; it provides the
correct isoscalar Z-graph contributions to leading order in $1/M$ for the
$NN$ system; and the systematic 3D formalism associated with it 
includes retardation effects without possessing the unphysical
singularities that generally are present in quasipotential theories.

For relativistic bound states, the 4D vertex functions at total
momentum $P$ obey 
\begin{equation}
\Gamma (P) = U(P) {\cal G}_0(P) \Gamma (P).
\label{eq:Gamma-4D}
\end{equation}
Equivalently, one may use the 3D vertex function that is
obtained with the use of
the ET propagator,
\begin{equation}
\Gamma _{\rm ET} (P) = U_1 (P) \langle{\cal G}_0(P) \rangle
\Gamma _{\rm ET}(P).
\label{eq:Gamma1-3D}
\end{equation}
Shortly, we shall need the relations of 4D vertex functions 
to the 3D ones, which is as follows,
\begin{equation}
\Gamma (P) = \Omega ^R (P) \Gamma_{\rm ET} (P),
\label{eq:OmegaRGamma}
\end{equation}
where $\Omega^R$ is defined by,
\begin{equation}
\Omega ^R (P)=  1 + [U(P) - U_1(P)] {\cal G}_0(P) \Omega^R (P) .
\label{eq:OmegaR}
\end{equation} 
A similar relation is needed for final states:
\begin{eqnarray}
\bar{\Gamma} (P) &=& \bar{\Gamma}_{\rm ET}(P) \Omega^L (P),
\label{eq:GammaOmegaL}\\
\Omega ^L (P)  &=& 1 + \Omega^L (P) {\cal G}_0 (P) [U(P) - U_1(P)].
\label{eq:OmegaL}
\end{eqnarray}

In the spirit underlying the use of an effective $NN$ interaction, the
3D interaction $U_1$ is truncated at ``lowest order'' within a
systematic expansion in powers of the coupling constant.  This
truncation violates Lorentz invariance, although the full 3D reduction
formalism is equivalent to the 4D formalism and thus respects Lorentz
invariance. For electromagnetic matrix elements, the absorption of the
virtual photon's momentum $Q$ causes the final state to have nonzero
three-momentum, even if the initial state did not. Thus, in general, a
dynamical boost is needed to obtain wave functions in frames where the
bound state has nonzero three-momentum.  In our 3D formalism, the dynamical
boost is embedded within the equations.  Wave functions corresponding
to nonzero three-momentum should be obtained straightforwardly by solving
the bound-state equation with the interaction appropriate to the
moving frame.  This differs from the interaction in the c.m. frame
because retardation corrections and Dirac spinors in the matrix
elements depend upon the total momentum.  The violation of Lorentz
invariance causes the total bound-state energy $E({\bf P})$ to differ from
$\sqrt{M_D^2 + {\bf P}^2}$.  For the deuteron, the effect is small and
it may be compensated by a simple renormalization of the interaction
that, in effect, approximately takes into account the terms omitted
when $U_1$ is truncated at lowest order.  Results for deuteron form
factors have been found in previous calculations to be essentially the same whether the interaction
is renormalized or not, thus indicating negligible sensitivity to the
truncation and the associated violation of Lorentz invariance.

In order to confront the predictions of this formalism
with electron-scattering data 
we must derive a 3D reduction of the electromagnetic current that is
consistent with the reduction of the scattering equation.
A clear 4D formalism for the current
follows from coupling photons everywhere in the Bethe-Salpeter
Green's function.
This leads to the following gauge-invariant 4D
result for the photon's interaction with the two-nucleon system:
\begin{equation}
{\cal A}_{{\rm BS},\mu} = \bar{\Gamma} (P') G^\gamma_{0,\mu} \Gamma
(P) + \bar{\Gamma} (P') G_0 (P') K^\gamma_{\mu} (Q) G_0 (P) \Gamma
(P),
\label{eq:GgammaBS}
\end{equation}
where $P' = P+Q$, $~G^\gamma_{0,\mu}$ is a five-point function for insertion
of a photon of momentum $Q$ and Lorentz index $\mu$ in the free propagator
$G_0$, and $K^\gamma_{\mu}$ is the five-point,  
two-nucleon irreducible $NN
\rightarrow NN \gamma$ amplitude found by coupling the photon to all
charged-particle lines inside the Bethe-Salpeter kernel $K$. 
A similar result holds for the 4D theory involving $G_C$,  
\begin{equation}
{\cal A}_{{\rm ET},\mu} = \bar{\Gamma} (P') {\cal G}^\gamma_{0,\mu}
\Gamma (P) + \bar{\Gamma} (P') {\cal G}_0 (P') U^\gamma_{\mu} (Q)
{\cal G}_0 (P) \Gamma (P),
\label{eq:GgammaET}
\end{equation}
where ${\cal G}^\gamma _{0,\mu} = G_{0,\mu}^\gamma + G^\gamma
_{C,\mu}$, and $U^\gamma_{\mu}$ is the result of coupling the photon
to all internal lines of the reduced interaction $U$.  In order to
develop a conserved current in the 4D theory based on
Eq.~(\ref{eq:T(U)4D}), we must construct $G^\gamma_{C,\mu}$, which is
the extra current required in order that ${\cal G}^\gamma_{0,\mu}$
satisfies a Ward-Takahashi identity involving ${\cal
G}_0$. Unfortunately, this is not straightforward, since in general
the variable $\kappa_2^0$ in $G_C$ depends on the three-momentum of
the intermediate state.  As we shall see in Section~\ref{sec-gi}, we
cannot simply use the standard fermion electromagnetic current,
$\gamma_\mu$, since this choice violates current conservation. In
Section~\ref{sec-gi} we construct the additional pieces of the current
that restore current conservation, thereby obtaining a conserved
current ${\cal G}_{0,\mu}^\gamma$ for use with vertex functions
obtained from Eq.~(\ref{eq:Gamma-4D}).

In this paper, we develop an equivalent 3D form of the
current matrix elements for the equal-time formalism,
based on an expansion of the formula,
\begin{equation}
{\cal A}_{{\rm ET},\mu} = \bar{\Gamma}_{\rm ET}(P') \langle
\Omega^L (P') {\cal G}^\gamma _{0,\mu} \Omega^R (P) \rangle \Gamma_{\rm ET}(P)
+ \bar{\Gamma}_{\rm ET} (P') \langle {\cal G}(P') 
U^\gamma_{\mu} {\cal G} (P) {\bf \rangle } \Gamma _{\rm ET}(P).
\label{eq:AmuET}
\end{equation}  
The resultant reduction is consistent with the reduction used to
obtain the 3D t-matrix of Eq.~(\ref{eq:3DET}) and the vertex function
of Eq.~(\ref{eq:Gamma1-3D}). Furthermore, this reduction preserves the
two-body electromagnetic Ward-Takahashi identities which are present
in the 4D theory. Quantities inside angle brackets are integrated over
time-components of relative momenta, thus reducing them to a 3D form.
This yields a consistent three-dimensional formalism that includes the
effects of relativity systematically, has the correct one-body limits,
and maintains current conservation. We then apply this machinery to
the calculation of electron-deuteron scattering in the impulse
approximation.

Previous calculations of electron-deuteron scattering by Hummel and
Tjon ~\cite{HT89,HT90,HT94} have used instant interactions and a form
of the ET current.  However, several approximations were employed,
such as the use of wave functions based on the 3D quasipotential
propagator of Blankenbecler-Sugar~\cite{BbS66} and
Logunov-Tavkhelidze~\cite{LT63}, approximate boost operators, and an
electromagnetic current which was conserved only in positive-energy
states.  Calculations of elastic electron-deuteron scattering also
were performed by Devine and Wallace using an instant interaction and
the ET propagator of Mandelzweig and Wallace, Eq.~(\ref{eq:aveG0GC})
with the choice of $\kappa_2^0$ given below that equation.~\cite{WD94}
In that work, a suitable conserved current was derived for use with
vertex functions that included negative-energy components.  In this
paper, we extend these previous analyses by use of our systematic 3D
formalism with the improved choice of $\kappa_2^0$ that is needed when
retardation effects are included.

It is easy to show that if the instant approximation is used to reduce
the four-dimensional equation, Eq.~(\ref{eq:T(U)4D}), to the
three-dimensional ET equation, a particularly simple three-dimensional
conserved current exists.  If we denote this current by ${\cal
G}_{{\rm inst},\mu}^\gamma$, then we may write the resulting
Ward-Takahashi identity as:
\begin{eqnarray}
Q^\mu {\cal G}_{{\rm inst},\mu}^\gamma(p_1,p_2;Q)%&=& q_2(\langle d_1(p_1)
%d_2(p_2) \rangle - \langle d_1(p_1) d_2(p_2 + Q) \rangle) +
%$(1 \leftrightarrow 2) \nonumber\\
&=&q_2[\langle {\cal G}_0 \rangle({\bf p}_1,{\bf p}_2;P^0) - \langle
{\cal G}_0 \rangle({\bf p}_1, {\bf p}_2 + {\bf Q};P^0 + Q^0)] 
\nonumber \\ &+& (1
\leftrightarrow 2),
\end{eqnarray}
where $(1 \leftrightarrow 2)$ indicates the part of the current
proportional to charge $q_1$, with momenta and labels of the two particles
interchanged.  
Consequently, if we construct the current matrix element 
\begin{equation}
{\cal A}_{{\rm inst},\mu}=\int \frac{d^3 p}{(2 \pi)^3} 
\bar{\Gamma}_{\rm inst}({\bf p};P+Q) 
{\cal G}_{{\rm inst},\mu}^\gamma({\bf p},{\bf P};Q)
\Gamma_{\rm inst}({\bf p};P),
\label{eq:emme}
\end{equation}
where $\Gamma_{\rm inst}$ is the vertex function obtained from the
bound-state equation corresponding to Eq.~(\ref{eq:Gamma1-3D}) with an
instant interaction chosen for $U_1$, then the result will be a
conserved impulse current.  A similar instant analysis was performed
by Devine and Wallace~\cite{WD95}.  Our instant calculations differ
from those of Ref.~\cite{WD95} in that a slightly different choice for
$\kappa^0_2$ is made than was made by Wallace and Mandelzweig. This
different choice for $\kappa_2^0$ is necessitated by the requirement
that the three-dimensional theory with retardations should not possess
any unphysical singularities when evaluated in a frame where the
two-body system has large momentum.

\begin{figure}[h]
\centerline{\BoxedEPSF{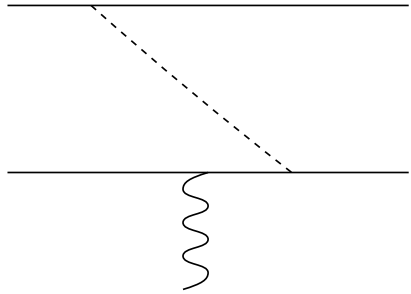 scaled 700}}
\vskip 5 mm
\caption{One example of a two-body current that is required in our formalism
in order to maintain gauge invariance.}
\label{fig-inflight}  
\end{figure}

A first step beyond the instant approximation may be obtained by
replacing the vertex $\Gamma_{\rm inst}$, which is calculated with
instant OBE interactions, with the vertex $\Gamma_{\rm ET}$,
calculated with the full retarded OBE interaction obtained in the
systematic ET formalism. In order to maintain current conservation our
approach demands that in making this change we should also replace
${\cal G}_{{\rm inst},\mu}^\gamma$ by a significantly more complicated
object, ${\cal G}_{0,\mu}^\gamma$. The resultant current matrix
element, which we denote ${\cal A}_{{\rm ET},\mu}$, differs from
Eq.~(\ref{eq:emme}) by replacement of the subscript {\rm inst} by the
subscript {\rm ET} in the vertex functions and replacement of ${\cal
G}_{{\rm inst},\mu}^\gamma$ by ${\cal G}_{0,\mu}^\gamma$.  This ${\cal
A}_{{\rm ET},\mu}$ would satisfy current conservation when
retardations are included provided that we also included the two-body
currents, such as the one depicted in Figure~\ref{fig-inflight}, that
become necessary because of retardation effects.  This work is
concerned primarily with displaying the formalism and performing
impulse approximation calculations that explore the effects of
retardation and negative-energy components in the vertex
functions. Therefore, we use the simpler current $\langle {\cal
G}_{{\rm inst},\mu}^\gamma \rangle$ with the ET vertex functions as
well as with the instant vertex functions.  For elastic
electron-deuteron scattering this is expected to be a very good
approximation to the use of the full ET current.  Calculations where
$\langle G_{C,\mu}^\gamma
\rangle$ is omitted from the instant current matrix element suggest
that the total contribution of $\langle G_{C,\mu}^\gamma \rangle$ to
the observables is small.  This is consistent with the interpretation
of $\langle G_{C,\mu}^\gamma\rangle$ as arising from the coupling of
the photon to internal charged lines in the crossed-box graph:
$\langle G_{C,\mu}^\gamma\rangle$ is a two-body current, and therefore
we would expect it {\it a priori} to be small at the
momentum-transfers under consideration here. Since $\langle
G_{C,\mu}^\gamma \rangle$ itself makes a small contribution to
observables, using a form of it that only approximately satisfies
current conservation is expected to have an even smaller effect on our
numerical results.  We expect that these effects, and the other
effects of retardations we have neglected in the current, will only
have a minor influence on observables. Here our main goal is to
examine the effect of retardation on the vertex functions, and the
role that negative-energy states play in the calculation, and so we do
not calculate these additional effects. Work is in progress to include
these extra terms of the current that are derived in Sec.~\ref{sec-gi}
in the calculation. We are also calculating meson-exchange currents,
such as $\rho \pi \gamma$ and $\omega \sigma \gamma$, which we {\it
do} expect to influence observables.

The first step in obtaining theoretical predictions for the
experimental observables in electron scattering on the deuteron is to
find the vertex functions $\Gamma_{\rm inst}$ and $\Gamma_{\rm ET}$.
We take the four-dimensional kernel $U$ in Eq.~(\ref{eq:T(U)4D}) to be
the sum of single-boson exchanges. The parameters for these exchanges
are taken from the Bonn-B OBE model, with the exception of the
$\sigma$-meson coupling, which we leave as a free parameter. From the
4D kernel, $U_{1,\rm OBE}$, an instant interaction, $U_{1,\rm inst}$,
is easily found, and a corresponding retarded interaction, $U_{1,\rm
ET}$, can be defined by the systematic procedure outlined above. These
two different interactions are then inserted into the bound-state
equation derived from Eq.~(\ref{eq:Gamma1-3D}). The $\sigma$-meson
coupling is adjusted so that the $NN$ bound-state pole in the
${}^3S_1-{}^3D_1$ channel appears at the deuteron mass. Once this is
done the deuteron vertex functions $\Gamma_{\rm inst}$ and
$\Gamma_{\rm ET}$ can be extracted. We calculate these two vertex
functions including the effects of negative-energy states and also in
the approximation where only positive-energy states contribute.

\begin{figure}[h]
\centerline{\BoxedEPSF{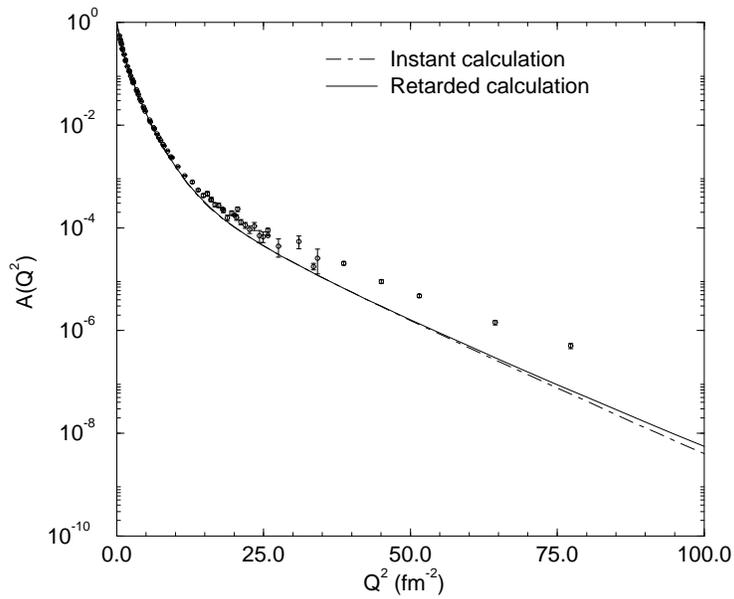 scaled 550}}
\vskip 5 mm
\caption{The form factor $A(Q^2)$ for the deuteron. The dash-dotted
line represents a calculation using a vertex function generated
using the instant interaction. Meanwhile the solid line is the result
obtained with the retarded ET vertex function. In both cases 
both positive and negative-energy sectors are included.}
\label{fig-Afull}  
\end{figure}

\begin{figure}[h]
\centerline{\BoxedEPSF{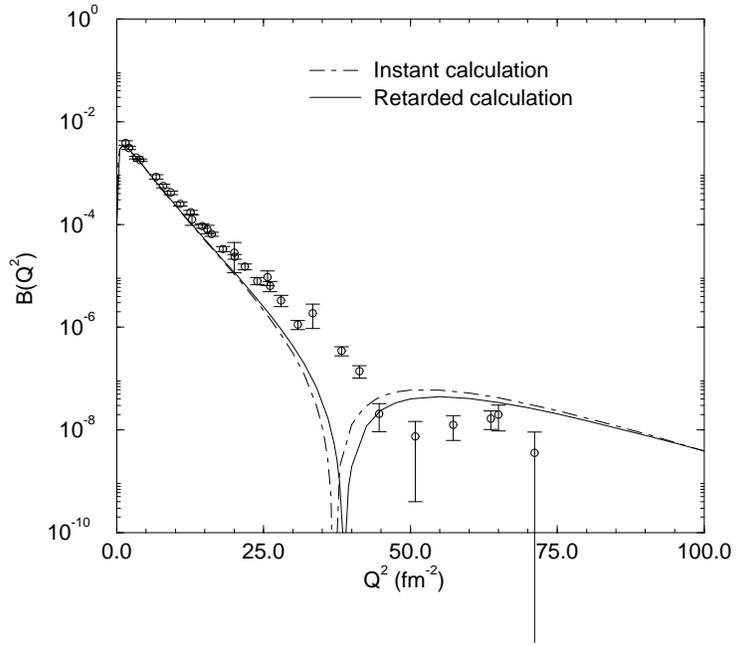 scaled 550}}
\vskip 5 mm
\caption{The form factor $B(Q^2)$ for the deuteron, legend as
in Fig.~\protect{\ref{fig-Afull}}.}
\label{fig-Bfull}  
\end{figure}

\begin{figure}[h]
\centerline{\BoxedEPSF{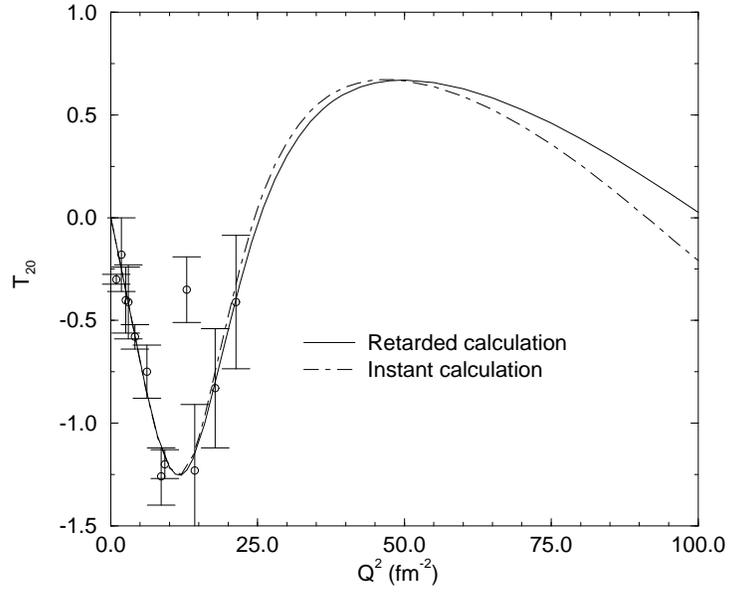 scaled 550}}
\vskip 5 mm
\caption{The tensor polarization $T_{20}(Q^2)$ in electron-deuteron
scattering, legend as in Fig.~\protect{\ref{fig-T20full}}.}
\label{fig-T20full}
\end{figure}

With vertex functions in hand and using the 3D current 
$\langle {\cal G}_{{\rm inst},\mu}^\gamma \rangle$,  we 
calculate the electron-deuteron scattering observables $A$, $B$, and
$T_{20}$. The results from such a calculation are shown in
Figs.~\ref{fig-Afull}, \ref{fig-Bfull} and
\ref{fig-T20full}.  We also show experimental data from 
Refs.~\cite{El69,Ar75,Si81,Cr85,Pl90} for $A$, from
Refs.~\cite{Si81,Cr85,Au85,Bo90} for $B$ and from
Refs.~\cite{Sc84,Dm85,Vo86,Gi90,Th91,Fe96} for $T_{20}$.  It should be
pointed out though, that a number of two-body effects, such as the
differences between ${\cal G}^\gamma_{{\rm inst},\mu}$ and ${\cal
G}^\gamma_{0,\mu}$ that are needed to restore current conservation, and
the usual $\rho \pi \gamma$ MEC contribution, should be added to our
calculations before they can be reliably compared to experimental
data. One of the most interesting features is the close similarity of
the results based on the ET vertex functions that include retardations
and instant vertex functions that do not.  We find that once the
$\sigma$ coupling is renormalized at $Q^2=0$ (as must be done to refit
the deuteron binding after incorporating the repulsive effects arising
from meson retardation), the deuteron properties in these two models
are remarkably similar.

\begin{figure}[h]
\centerline{\BoxedEPSF{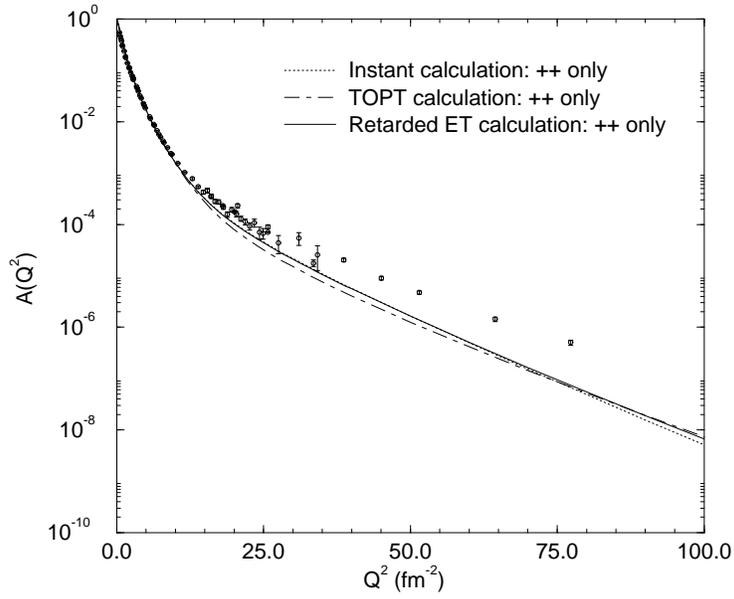 scaled 550}}
\vskip 5 mm
\caption{The form factor $A(Q^2)$ for the deuteron, as predicted by
three different $NN$ models. The dotted curve is the instant
calculation, the dash-dotted curve is the calculation using the TOPT
interaction, and the solid curve is the result using a retarded ET
interaction. All calculations are done with the contributions of
negative-energy states dropped.}
\label{fig-plusplusonlyA}
\end{figure}

\begin{figure}[h]
\centerline{\BoxedEPSF{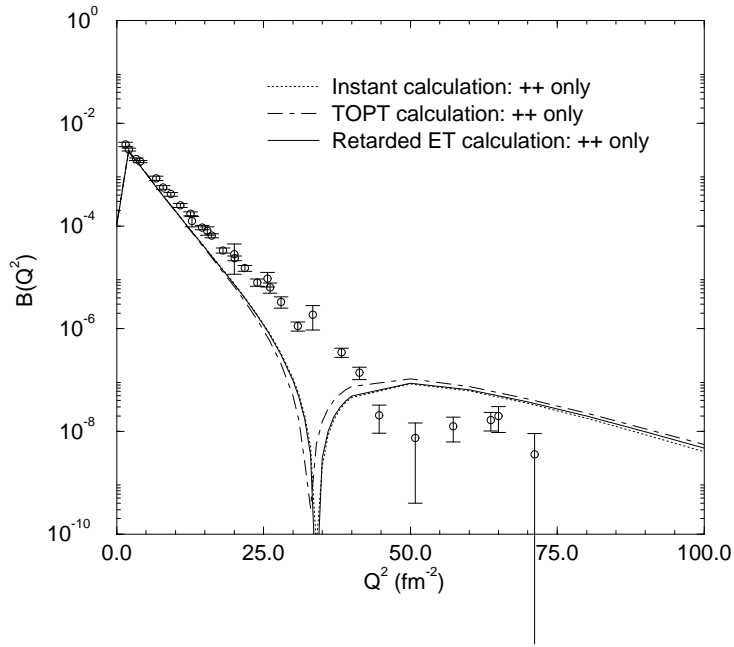 scaled 550}}
\vskip 5 mm
\caption{The form factor $B(Q^2)$ for the deuteron, legend as in
Fig.~\protect{\ref{fig-plusplusonlyA}}. Again, all calculations are done
in the $++$ only approximation.}
\label{fig-plusplusonlyB}  
\end{figure}

\begin{figure}[h]
\centerline{\BoxedEPSF{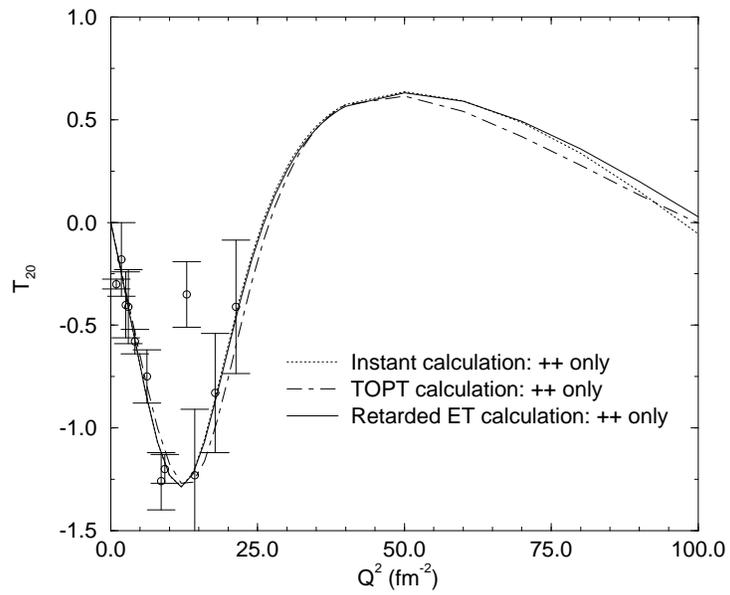 scaled 550}}
\vskip 5 mm
\caption{The tensor polarization $T_{20}(Q^2)$ in electron-deuteron scattering,
legend as in Fig.~\protect{\ref{fig-plusplusonlyA}}. All calculations are
done using only $++$ states.}
\label{fig-plusplusonlyT20}  
\end{figure}

For comparison, in Figs.~\ref{fig-plusplusonlyA}, \ref{fig-plusplusonlyB} and
\ref{fig-plusplusonlyT20} we display calculations for $A$, $B$ and $T_{20}$, 
where the effects of negative-energy states are removed.  We see that
the inclusion of these states in the calculation makes little
difference to any of the observables. However, comparing the different
positions of the minima in Figs.~\ref{fig-Bfull} and
\ref{fig-plusplusonlyB} we see that including the negative-energy
states in the calculation does shift the minimum in $B$ to somewhat
larger $Q^2$. A similar effect was observed by van Orden {\it et
al.}~\cite{vO95} in calculations of electron-deuteron scattering using
the spectator formalism. However, note that here, in contradistinction
to the results of Ref.~\cite{vO95}, the inclusion of negative-energy
states does {\it not} bring the impulse approximation calculation into
agreement with the data.

The fact that negative-energy states seem to have a smaller effect on
observables in the ET analysis than in the spectator analysis of van
Orden et al.~\cite{vO95} is somewhat surprising.  As pointed out
above, the ET propagator has twice the negative-energy state
propagation amplitude of the Gross propagator. Thus, other differences
between the ET and spectator models, not just differences in the role
of negative-energy states in the two approaches, appear to be
responsible for the differing results for $B(Q^2)$. The fact that
Ref.~\cite{vO95}'s calculation essentially agrees with the
experimental data for $B(Q^2)$ is not solely attributable to the Gross
formalism's treatment of negative-energy states, since our results
show that negative-energy states have a much smaller effect.

We also note that the inclusion of retardation moves the minimum in
$B$ a little higher in $Q^2$ but otherwise has little observable
effect. Again, this additional effect is not enough to bring the
predictions of our model into line with the experimental data. 

Finally, examining the tensor polarization $T_{20}$ we see that all
the different models produce results which are very similar. This
suggests that this observable is fairly insensitive to dynamical
details of the deuteron model, at least up to $Q^2=4 \,\, \rm{GeV}^2$.

We find that elimination of the approximations used by Hummel and Tjon
produces only minor changes for the experimentally-measured quantities
$A$, $B$, and $T_{20}$, although the precise location of the minimum
in the magnetic form factor $B$ does change when meson retardations
are included in the calculation. Nevertheless, there is little
improvement in the agreement of the impulse-approximation calculations
with the experimental data for $B$. 

Comparing our results with those of a non-relativistic impulse
approximation calculation of the same observables (see, for instance,
Ref.~\cite{Ar80}) strengthens the conclusion of previous authors that
neither the consideration of relativistic kinematics for the nucleons
and mesons nor the inclusion of negative-energy state effects improves
the agreement with the experimental data. Of course, by definition,
such a non-relativistic impulse approximation calculation neglects
both relativistic effects and two-body current contributions.  Indeed,
we see here that, as already found by Arnold {\it et al.}~\cite{Ar80}
and Zuilhof and Tjon~\cite{ZT81}, including relativistic effects
actually {\it worsens} the agreement with the experimental data. The
inclusion of these effects to all orders in a $p/M$ expansion, as done
here and in Refs.~\cite{Ar80,ZT81} does {\it not} lead to a small
correction which brings the theory into closer agreement with the
experimental data. This suggests that the comparative success of a
simple non-relativistic impulse approximation calculation is
fortuitous. Dynamical mechanisms beyond the impulse approximation
appear to play a more important role in this reaction than one would
conclude from the non-relativistic
calculation. Figures~\ref{fig-Afull}, \ref{fig-Bfull}, and
\ref{fig-T20full} imply that once $Q^2$ gets above about 0.5
$\rm{GeV}^2$ two-body current contributions become important. One
example of such a two-body current would be meson-exchange current
contributions, but it is also possible that other two-body currents
are dynamically important in this regime.

The rest of this paper is devoted to a detailed description of our
formalism, and to the explanation of technical details pertaining to
our calculation that were omitted in the brief sketch of the
calculations given here. In
Section~\ref{sec-eikonal} we present our modified four-dimensional
equation for $G_C$ with the correct one-body limit and incorporating a
form of the eikonal approximation which is an improvement over that
proposed in Ref.~\cite{PW96}.  The new choice of the parameter
$\kappa^0_2$ has been found to be necessary, when retardations
are included in the ET interaction, in order to avoid unphysical
singularities when ${\bf P}$ is large.  Our formalism for reductions to three
dimensions is then used to determine the lowest-order 3D interaction
$U_1$.  In Section~\ref{sec-potentials} we explain the various
interactions that are used in calculations of deuteron wave functions.
These can be divided into two classes: instant interactions, and
ET interactions that include meson retardation.  Within either of these
classes, versions of the interactions are constructed that do and do not
include the effects of negative-energy states, in order to display the
role played by such components of the deuteron wave function.
Section~\ref{sec-gi} discusses our 3D reduction of the electromagnetic
current that maintains current conservation, including a detailed
discussion of the way in which the gauging of the propagator $G_C$ in
our modified four-dimensional equation is performed. The ET current
${\cal G}^\gamma_{0,\mu}$ is derived and we discuss its connection to the
simpler current ${\cal G}^\gamma_{{\rm inst},\mu}$ that is used in the present
calculations.  Section ~\ref{sec-results} presents 
details concerning the application of this formalism to the calculation of
electromagnetic observables using the wave functions computed as
discussed in
Section~\ref{sec-potentials}.  Conclusions are presented in
Section~\ref{sec-conclusion}.

\section {Bound-state equations with correct one-body limits}

\label{sec-eikonal}

\subsection{A four-dimensional equation}

As outlined in the Introduction, a simple four-dimensional equation
may be obtained by employing a form of the eikonal approximation to
reorganize the generalized ladder Bethe-Salpeter kernel, based upon
the fact that we may always write the kernel $K$ in the iterative
form, Eq.~(\ref{eq:U-4D}).  The question is: what are good choices for
$G_C$ and $U$? To answer this question, consider the lowest order (in
coupling constant) parts of the kernels $K$ and $U$.  Obviously, the
second-order pieces must be the same, i.e.,
\begin{equation}
U^{(2)} = K^{(2)}.  
\end{equation}
Meanwhile, at fourth order, we obtain
\begin{equation}
K^{(4)} = U^{(4)} + U^{(2)} G_C U^{(2)}.  
\label{eq:U4D}
\end{equation}
Thus, in order that the expansion for $U$ can be truncated at second
order without losing the one-body limit as $m_2 \rightarrow \infty$,
the choice of $G_C$ must be such that the last term captures the part
of $K^{(4)}$ that is non-zero in that limit. Moreover, if $G_C$ is
chosen in this way, the one-body limit will then be correct no matter 
the order at which 
the expansion of $U$ is truncated. 

\begin{figure}
\centerline{\BoxedEPSF{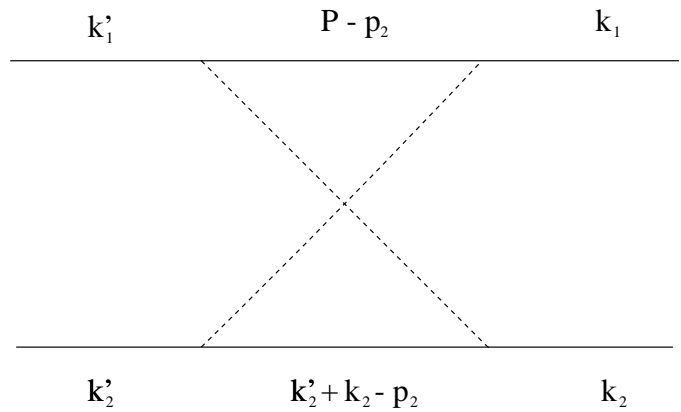 scaled 500}}
\vskip 5 mm
\caption{The crossed-box graph, showing the momentum labels used
in the text.}
\label{fig-crossedbox}
\end{figure}

Consider the fourth-order contribution to $K$, i.e., the crossed-box
graph depicted (with momentum labels) in Fig.~\ref{fig-crossedbox}.
In order to express $K^{(4)}$ in the form (\ref{eq:U4D}), it is
necessary to commute vertices until they appear in the same order as
in $K^{(2)} G_C K^{(2)}$. In doing this commutator terms are collected
in $U^{(4)}$.  Once this is done, the part of the graph that takes on
an iterative form is expressed as
\begin{eqnarray}
K_{\rm iter}^{(4)}(k_1',k_2';k_1,k_2)=i \int \frac{d^4 p_2}{(2 \pi)^4}
K^{(2)}(k_2'-p_2) d_1(P-p_2) d_2(k_2 + k_2' - p_2) K^{(2)}(p_2 - k_2).
\label{eq:3.6}
\end{eqnarray}
Here $K^{(2)}$ is the ladder Bethe-Salpeter kernel, 
$P$ is the conserved, total four-momentum: 
\begin{equation}
P=k_1 + k_2=k_1' + k_2',
\end{equation}
and the propagators $d_i$ are given by, 
\begin{equation}
d_i(p_i)=\frac{\Lambda_i^+({\bf p}_i)}{p^0_i - \epsilon_i ({\bf p}_i) + i \eta}
- \frac{\Lambda_i^-({\bf p}_i)}{p^0_i + \epsilon_i ({\bf p}_i) - i \eta},
\label{eq:di}
\end{equation}
where
\begin{eqnarray}
\epsilon_i({\bf p}_i)&=&({\bf p}_i^2 + m_i^2)^{1/2},\\ \nonumber  \\
\Lambda^{\pm}_i({\bf p}_i)&=&\left \{ 
\begin{array}{ll}
\frac{1}{2 \epsilon_i({\bf p}
_i)}, & \mbox{for spin-zero particles,}\\
\frac{\pm \epsilon_i ({\bf p}_i) \gamma^0 - {\bf \gamma}_i\cdot {\bf p}_i+m_i}
{2 \epsilon_i({\bf p}_i)}, & \mbox{for spin-half particles,}
\end{array} \right  \}
\label{eq:Lambda}
\end{eqnarray}
and $\eta$ is a positive infinitesimal.  

In Eq.~(\ref{eq:3.6}) the argument of the function $d_2$ may be
rewritten:
\begin{equation}
(k_2^0 + k_2^{\prime 0} - p_2^0,{\bf k}_2 + {\bf k}_2' - {\bf p}_2).
\end{equation}
Suppose that the zeroth component of the momentum of particle 2 is
large. In particular, this will be the case if $m_2 \gg m_1$ (one-body
limit).  In this case particle two's intermediate and final-state
three-momentum will be largely unaffected by the presence of particle
one, and so we may approximate these components as unchanging,
\begin{equation}
{\bf k}_2 + {\bf k}_2' \approx 2 {\bf p}_2.
\label{eq:eikapprox}
\end{equation}
This is the eikonal approximation, and it rests upon forward scattering
being the dominant mechanism of interaction.  
Indeed, making the replacement (\ref{eq:eikapprox}) in
(\ref{eq:3.6}) will not affect the value of $K_{\rm iter}^{(4)}$ in the limit $m_2 
\rightarrow \infty$.

This argument shows that $K_{\rm iter}^{(4)}$ may be approximately 
rewritten as:
\begin{eqnarray}
K_{\rm iter}^{(4)}(k_2',k_2;P) \approx i \int \frac{d^4 p_2}{(2 \pi)^4}
K^{(2)}(k_2' - p_2) d_1(P-p_2) d_2(2 \kappa_2 - p_2) K^{(2)}(p_2-k_2),
\label{eq:K4it}
\end{eqnarray}
with the four-component object $\kappa_2$ defined by:
\begin{equation}
\kappa_2=\left(\frac{k_2^0 + {k_2'}^0}{2},{\bf p}_2\right).
\label{eq:kappa2}
\end{equation}
In operator notation
\begin{equation}
K_{\rm iter}^{(4)} \approx K^{(2)} G_C K^{(2)}.
\label{eq:Kit}
\end{equation}
Now we are in a position to answer the question of how to 
choose $G_C$. Equation~(\ref{eq:K4it}) shows that the choice:
\begin{equation}
G_C(p;P) =i d_1(\nu_1 P + p) d_2(2 \kappa_2 - \nu_2 P + p),
\label{eq:GC2}
\end{equation}
which we write symbolically as,
\begin{equation}
G_C = i d_1 d_2^c,
\label{eq:GC3}
\end{equation}
will allow $U$ to be truncated at second order while still
yielding a four-dimensional equation with the correct one-body limit.
Here it is understood that when writing $G_C$ as a function of 
the four four-momenta $p_1'$, $p_2'$, $p_1$, and $p_2$, we have
\begin{equation}
G_C(p_1',p_2';p_1,p_2)=i (2 \pi)^8 \delta^{(4)}(p_1'-p_1) 
\delta^{(4)}(p_2'-p_2) d_1(p_1) d_2(2 \kappa_2 - p_2).
\end{equation}
On the other hand, in writing Eq.~(\ref{eq:GC2}) we have expressed the result
in terms of center-of-mass and relative four-momenta:
\begin{eqnarray}
p_1=\nu_1 P + p&,& \qquad \qquad p_2=\nu_2 P - p;\\ \nonumber \\
\nu_1=\frac{P^2 + m_1^2 - m_2^2}{2 P^2}&,& \qquad \qquad
\nu_2=\frac{P^2 - m_1^2 + m_2^2}{2 P^2}.
\end{eqnarray}
The propagator $d_2^c$ defined in (\ref{eq:GC2}) depends on the zeroth
components of the external momenta through the zeroth component of
$\kappa_2$. Thus the use of operator notation in Eq.~(\ref{eq:Kit})
anticipates a suitable choice for these zeroth components of external
momenta.

In the case that particle two is extremely massive, it will not be very
far off its mass shell.  The total energy $P^0$ of the system will be
approximately the energy of particle two, and thus 
we choose
\begin{equation}
\kappa_2^0=\nu_2 P^0.
\label{eq:nu2P0}
\end{equation}
This choice was used in Ref.~\cite{PW96} because it leads to an
integral equation that satisfies both the criteria needed to obtain
the correct one-body limit $m_2 \rightarrow \infty$.  It also is
suitable for equal-mass particles in the center-of-mass~frame.  However, in
frames where the bound state has nonzero total momentum, the choice
leads to an interaction that possesses unphysical singularities.

A consideration of the singularities of the interaction when the particles have
equal masses suggests that the particles should be 
equally far off their mass shells in the intermediate state.  Hence, in that
case we choose
\begin{equation}
\kappa_2^0=\frac{1}{2}\left(P^0 - \epsilon_1 + \epsilon_2\right),
\label{eq:equalmasskappa20}
\end{equation}
where, in an arbitrary frame,
\begin{equation}
\epsilon_1=\epsilon_1({\bf P}/2 + {\bf p}); \qquad 
\epsilon_2=\epsilon_2({\bf P}/2 - {\bf p}).
\end{equation}
This choice is consistent with the eikonal approximation and differs
from that of Eq.~(\ref{eq:nu2P0}) only when ${\bf P} \neq 0$.  Note
that our new choice for $\kappa_2^0$ depends not only on the external
variables $P^0$ and ${\bf P}$, but also on the internal three-momentum
${\bf p}$.  This feature complicates the construction of a conserved
current.  

Once $U$ and the $G_C$ of Eq.~(\ref{eq:GC2}) are chosen, this
defines an ``improved'' ladder BSE, which, in the two-body
center-of-mass frame takes the form,
\begin{eqnarray}
\Gamma({p'}^0,{\bf p}';s)&=&i \int \frac{d^4 p}{(2 \pi)^4}
U(p',p) d_1(\nu_1 P^0 +p^0,{\bf p})\nonumber\\ &&
\left[d_2(\nu_2 P^0-p^0,-{\bf p}) + d_2(2 \kappa_2^0 -
\nu_2 P^0 + p^0,-{\bf p}) \right] \Gamma(p^0,{\bf p};s).
\label{eq:3.19}
\end{eqnarray}
Because the choice of $\kappa_2^0$ employed here agrees with
that of Ref.~\cite{PW96} in the limit $m_2 \rightarrow \infty$ the
proof given there now suffices to show that this equation becomes the
appropriate one-body equation in the limit $m_2 \rightarrow \infty$.
For equal mass particles, $\kappa_2^0$ is chosen as per
Eq.~(\ref{eq:equalmasskappa20}).

We make the choice (\ref{eq:equalmasskappa20}) because it eliminates
singularities that afflict the choice (\ref{eq:nu2P0}). As observed in
Ref.~\cite{PW96} if the choice (\ref{eq:nu2P0}) is made for
$\kappa_2^0$ then the equation (\ref{eq:3.19}) develops a cut
beginning at
\begin{equation}
2 \nu_2 P^0=2 m_2 + \mu.
\end{equation}
If we attempt to solve the integral equation (\ref{eq:3.19}) in a frame 
moving with large enough three-momentum ${\bf P}$, then $P^0 =
\sqrt{M^2_d + {\bf P}^2 }$ will become big enough to encounter
this singularity. However, with the choice (\ref{eq:equalmasskappa20})
the interaction does not have this unphysical singularity. 

The above arguments pertain to the one-body limit in which $m_2 \rightarrow
\infty$.  Clearly we can interchange the labels on particles one and
two in order to obtain a propagator $i d_1^c d_2$ that is motivated
by a consideration of the limit $m_1 \rightarrow \infty$. In the propagator
$d_1^c$ we choose, for the equal-mass case
\begin{equation}
\kappa_1^0=\frac{1}{2} (P^0 + \epsilon_1 - \epsilon_2).
\label{eq:equalmasskappa10}
\end{equation}

In the case $m_1=m_2$, writing the propagator as in Eq.~(\ref{eq:GC3})
treats the two particles asymmetrically.  If they are identical
particles this will lead to violations of the identical-particle
statistics.  To avoid this, we symmetrize the propagator $G_C$, by
choosing, instead of $i d_1 d_2^c$, the form
\begin{eqnarray}
{\cal G}_0 &\equiv& G_0 + G_C \nonumber \\
  &=& \frac{i}{2} (d_1 + d_1^c)(d_2 + d_2^c).
\label{eq:G0plusGC}
\end{eqnarray}
If the mass of particle $n$ is taken to infinity we use the choice
$\kappa_n^0=P^0$ in $d_n^c$, while the $\kappa^0$ variable associated
with the light particle is chosen to be zero. This yields the one-body
propagator for the light particle, multiplied by a delta function in
the relative momentum, and a projection operator onto the
positive-energy states of particle $n$. Meanwhile, the delta function
guarantees that the interaction will be static, and so both conditions
necessary for the correct one-body limit to be present are
satisfied. However, note that this formal property does not apply to
the case of practical interest for this work, $m_1=m_2$, since then 
$\kappa_1^0$ and $\kappa_2^0$ are defined by Eqs.~(\ref{eq:equalmasskappa10}) and
(\ref{eq:equalmasskappa20}).

 Symbolically we now write our ``improved'' ladder BSE as
\begin{equation}
\Gamma=U {\cal G}_0 \Gamma,
\label{eq:4DET}
\end{equation}
where here, and throughout this work, ${\cal G}_0$ is
defined by Eq.~(\ref{eq:G0plusGC}).

We stress that what has been done here is to take certain pieces of
the Bethe-Salpeter kernel $K$ and rewrite them as $K^{(2)} G_C
K^{(2)}$, $K^{(2)} G_C K^{(2)} G_C K^{(2)}$, etc. Consequently,
Eq.~(\ref{eq:4DET}) is equivalent to a Bethe-Salpeter equation in
which graphs other than one-meson exchange are approximately included
in the kernel. Thus we expect that the solution of this equation may
provide a better description of the dynamics of two-particle systems
than the ladder BSE amplitude. Numerical calculations in a scalar
field theory appear to support this~\cite{PW96}.

\subsection{Summary of the reduction to three dimensions}

The 3D reduction that was outlined in the Introduction may now be
applied to the 4D equation (\ref{eq:4DET}).  This yields the following
3D equation for the bound-state vertex function, $\Gamma_{\rm
ET}$,
\begin{equation}
\Gamma_{\rm ET}=U_1 \langle {\cal G}_0 \rangle \Gamma_{\rm ET},
\label{eq:ET}
\end{equation}
with $\langle {\cal G}_0 \rangle$ the three-dimensional propagator of
Eq.~(\ref{eq:aveG0GC}), and the interaction $U_1$ defined as the
solution of the following coupled equations,
\begin{equation}
U_1 = \langle {\cal G}_0 \rangle ^{-1} \langle {\cal G}_0 U {\cal G} \rangle
\langle {\cal G}_0 \rangle ^{-1} ,
\label{eq:K1.2}
\end{equation}
which is three-dimensional, and 
\begin{equation}
{\cal G} = {\cal G}_0 + {\cal G}_0 (U - U_1) {\cal G} .
\label{eq:calG0-GC}
\end{equation}
Note once more that the solution for $U_1$ is found by demanding that
$\langle {\cal G} \rangle = \langle {\cal G}_0 \rangle$.
Equations~(\ref{eq:ET}) and (\ref{eq:K1.2}) are exactly equivalent to
the 4D equation (\ref{eq:4DET}).  Inserting the choice
(\ref{eq:equalmasskappa20}) for $\kappa_2^0$ and specializing to the
case of equal-mass particles gives
\begin{equation}
\langle {\cal G}_0(P) \rangle=\frac{\Lambda_1^+ \Lambda_2^+}{{P^0} - \epsilon_1
- \epsilon_2} - \frac{\Lambda_1^+ \Lambda_2^-}{2 \epsilon_2} -
\frac{\Lambda_1^- \Lambda_2^+}{2\epsilon_1} 
- \frac{\Lambda_1^- \Lambda_2^-}{{P^0} + \epsilon_1 + \epsilon_2},
\label{eq:ET-prop}
\end{equation}
which is a restatement of Eq.~(\ref{eq:aveG0GC}).
We note in passing that this is exactly the same expression as that found when
the simpler expression $i \langle d_1 (d_2 + d_2^c) \rangle$ is calculated. 

By  using Eq.~(\ref{eq:ET}), rather  then  Eq.~(\ref{eq:Salpeter}), we
not only have a consistent  3D reduction but we approximately include
the  crossed-ladder graphs, particularly the ``Z-graph'' contributions
where one nucleon is in a positive-energy state and the  other is in a
negative-energy state.

For equal-mass particles in the center-of-mass frame,
Eq.~(\ref{eq:ET}) agrees with the quasipotential equations of Wallace
and Mandelzweig~\cite{WM89}.
However, in a quasipotential approach the
interaction in other frames should be obtained by boosting via a
dynamical equation.  This leads to unphysical singularities.  The
systematic ET formalism avoids these.  In essence, the ET formalism
with a truncated interaction emphasizes the elimination of unphysical
singularities over the strict enforcement of Lorentz invariance.

The second-order 3D interaction $U_1^{(2)}$ is given by:
\begin{equation}
\langle {\cal G}_0 \rangle U_1^{(2)}\langle {\cal G}_0 \rangle=
\langle {\cal G}_0 K^{(2)} {\cal G}_0 \rangle \equiv {\cal A}.
\end{equation}
An explicit form for ${\cal A}$ can be computed for the case of OBE
interactions of the form
\begin{equation}
K^{(2)}(q) = \sum _{n} \frac{g_n^2 V_n(1) V_n(2)}{q^2 - \mu_n^2} 
\end{equation}
where $V_n(i)$ denotes the appropriate vertex operator for the interaction
of the $n$th meson with nucleon $i$. If we write:
\begin{equation}
{\cal A}=\sum_n \sum_{\rho_1 \rho_1' \rho_2 \rho_2'} 
\frac{\Lambda^{\rho_1' }_1 \Lambda^{\rho_2' }_2}{e_1' + e_2'} 
A_n(\rho_1' \rho_2' \leftarrow \rho_1 \rho_2)V_n(1) V_n(2) 
\frac{\Lambda^{\rho_1}_1 \Lambda^{\rho_2}_2}{e_1 + e_2},
\label{eq:4.11}
\end{equation}
with $e_i=\rho_i \nu_i \kappa_i^0 - \epsilon_i$, $e_i'=\rho_i' \nu_i
{\kappa_i'}^0 - \epsilon_i'$; then, for the exchange of meson $n$, the
factors $A_n(\rho_1' \rho_2' \leftarrow \rho_1 \rho_2)$ can be written
in any rho-spin channel as
\begin{eqnarray} 
A_n({\bf p}',{\bf p};P^0, {\bf P})(\rho_1' \rho_2' \leftarrow \rho_1
\rho_2)=\frac{g_n^2 }{8 \omega_n}
\left[ F (\omega_n + q^0) + F(\omega_n - q^0) \right] , 
\label{eq:U1symm}
\end{eqnarray}
where 
\begin{eqnarray}
&& F(\omega) \equiv 
\frac{1}{e_1 + e_2' - \omega}
+ \frac{1}{e_1' + e_2 - \omega} 
+ \frac{1}{e_1 + e_1' - \omega}
+ \frac{1}{e_2 + e_2' - \omega} 
\nonumber\\
&& \quad \quad +  \frac{1}{e_1 + e_2' - \omega} 
\left( \frac{e_1 + e_2}{e_2 + e_2' - \omega}\right)
+  \frac{1}{e_1' + e_2 - \omega}
\left( \frac{e_1 + e_2}{e_1 + e_1' - \omega}\right)\nonumber\\
&& \quad \quad +  \left( \frac{e_1' + e_2'}{e_1 + e_1' - \omega}\right) 
\frac{1}{e_1 + e_2' - \omega}
+ \left( \frac{e_1' + e_2'}{e_2 + e_2' - \omega}\right) 
\frac{1}{e_1' + e_2 - \omega}\nonumber\\
&& \quad \quad + \left. \frac{e_1' + e_2'}{e_2 + e_2' - \omega} 
\left(\frac{1}{e_1 + e_2' - \omega} + \frac{1}{e_1' + e_2 - \omega} \right) 
\frac{e_1 + e_2}{e_1 + e_1' - \omega} \right.  ,
\end{eqnarray}
and $q_0={\kappa_1'}^0 - \kappa_1^0$. The key
difference here from the calculation of Ref.~\cite{PW96} is that here the
parameter $\kappa_i^0$ varies with the intermediate-state momentum
under consideration. For equal-mass particles
\begin{eqnarray}
\kappa_1^0&=&\frac{1}{2} \left(P^0 + \epsilon_1 - \epsilon_2\right), \qquad
\kappa_2^0=\frac{1}{2} \left(P^0 - \epsilon_1 + \epsilon_2\right);\\
{\kappa_1'}^0&=&\frac{1}{2} \left(P^0 + \epsilon_1' -
\epsilon_2'\right), \qquad {\kappa_2'}^0=\frac{1}{2} \left(P^0 -
\epsilon_1' + \epsilon_2'\right).
\end{eqnarray}
The expression for ${\cal A}$ is easily converted into a result for $U_1^{(2)}$
using the inverse of the propagator $\langle {\cal G}_0 \rangle$.
The result is expressed in terms of matrix elements between
initial and final states of any $\rho$-spin as follows, 
\begin{eqnarray}
U_{1,{\rm ET}}^{\rho_1' \rho_2' ,\rho_1 \rho_2}({\bf p}',{\bf p};P^0,
{\bf P})= && \sum_n \left[\bar{u}^{\rho_1'}(\rho_1'{\bf p}_1') V_n(1)
u^{\rho_1}(\rho_1{\bf p}_1)\right]
\left[\bar{u}^{\rho_2'}(\rho_2'{\bf p}_2') V_n(2) u^{\rho_2}(\rho_2{\bf
p}_2)\right] \nonumber \\ && \times    
 A_n({\bf p}',{\bf p};P^0, 
{\bf P})(\rho_1' \rho_2' \leftarrow \rho_1 \rho_2)  
\label{eq:U1ET}
\end{eqnarray}
This is the full ET interaction including retardation effects.  
If one eliminates the retardation effects, the result is the instant
interaction defined by,
\begin{equation}
U_{1,{\rm inst}}^{\rho_1' \rho_2' ,\rho_1 \rho_2}({\bf p}',{\bf p};P^0, {\bf
P})= \sum_n \left[\bar{u}^{\rho_1'}(\rho_1'{\bf p}_1') V_n(1) 
 u^{\rho_1}(\rho_1{\bf p}_1)\right] \left[ \bar{u}^{\rho_2'}(\rho_2'{\bf p}_2')
V_n(2) u^{\rho_2}(\rho_2{\bf p}_2)\right] \left( \frac{-g_n^2}{\omega_n^2}\right). 
\label{eq:U1inst}
\end{equation}

Expanding the vertex functions in terms of
Dirac spinors,
\begin{equation}
\Gamma_{\rm ET} ({\bf p}, {\bf P}) = \gamma_1^0 \gamma_2^0
\sum _{\rho_1 \rho_2} u^{\rho_1}(\rho_1 {\bf p}_1) 
u^{\rho_2} (\rho_2 {\bf p}_2) \Gamma _{\rm ET}^{\rho_1 \rho_2} ({\bf
p}, {\bf P}),
\end{equation}
leads to the coupled equations that we solve,
\begin{equation}
\Gamma ^{\rho_1,\rho_2}_{\rm ET} ({\bf p},{\bf P}) = \sum_{\rho_1' \rho_2'}
\int \frac{d^3 p'}{(2 \pi)^3} 
U_1^{\rho_1 \rho_2, \rho_1' \rho_2'} ({\bf p}, {\bf p}';P^0,{\bf P}) 
\langle {\cal G}_0 \rangle^{\rho_1' \rho_2'}({\bf p}', {\bf P})  
\Gamma_{\rm ET}^{\rho_1' \rho_2'} ({\bf p}', {\bf P}).
\label{eq:ET-Gamma}
\end{equation}
Here the propagator factors $\langle {\cal G}_0\rangle^{\rho_1' \rho_2'}$ are
$\langle {\cal G}_0 \rangle^{++} = 1/(P^0 - \epsilon_1' -
\epsilon_2')$, $\langle {\cal G}_0 \rangle^{+-} = 1/ 2 \epsilon_2'$,  
$\langle {\cal G}_0\rangle^{-+} = 1/ 2 \epsilon_1' $ and $\langle
{\cal G}_0 \rangle^{--} = -1/(P^0 + \epsilon_1' + \epsilon_2' )$.

\begin{figure}[h]
\centerline{\BoxedEPSF{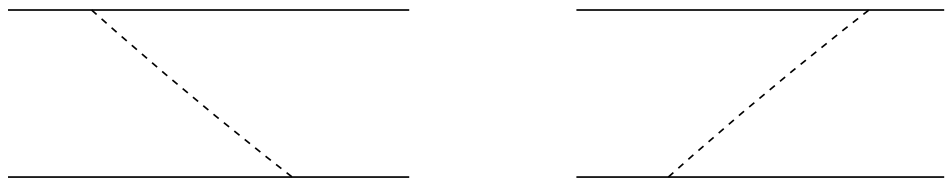 scaled 700}}
\vskip 5 mm
\caption{The two time-ordered perturbation theory graphs for
one-pion exchange.}
\label{fig-OPE}  
\end{figure}

Two special cases of particular interest are the matrix element of 
the ET interaction between positive-energy
spinors, as follows,  
\begin{eqnarray}
&&  U_{1,{\rm ET}}^{++,++}({\bf p}',{\bf
p};P^0, {\bf P}) = \sum_n \left[\bar{u}^+({\bf p}_1') V_n(1) u^+({\bf p}_1)\right]
\left[\bar{u}^+({\bf p}_2') V_n(2) u^+({\bf p}_2)\right]
\nonumber \\ && \times \frac{g_n^2}{2 \omega_n} 
\left[ \frac{1}{P^0 - \epsilon_1' - \epsilon_2
- \omega_n} + \frac{1}{2} \frac{P^0 - \epsilon_1 - \epsilon_2}{(P^0 -
\epsilon_1' - \epsilon_2 - \omega_n)^2} \right. \nonumber\\ && \qquad \qquad  
\left. + \frac{1}{2} \frac{P^0 - \epsilon_1' -
\epsilon_2'}{(P^0 - \epsilon_1' - \epsilon_2 - \omega_n)^2} 
+ \frac{1}{2} \frac{(P^0 - \epsilon_1' - \epsilon_2')(P^0 - \epsilon_1 -
\epsilon_2)} {(P^0 - \epsilon_1' - \epsilon_2 - \omega_n)^3} + (1
\leftrightarrow 2)\right],
\label{eq:U1ET++}
\end{eqnarray}
and the matrix element in positive-energy states omitting the
effects of  $G_C$, 
\begin{eqnarray}
 U_{1,{\rm TOPT}}^{++,++}({\bf p}',{\bf
p};P^0, {\bf P})&& = \sum_n \left[\bar{u}^+({\bf p}_1') V_n(1) u^+({\bf p}_1)\right]
\left[\bar{u}^+({\bf p}_2') V_n(2) u^+({\bf p}_2)\right]
\frac{g_n^2}{2 \omega_n} \nonumber \\
&& \times \left[ \frac{1}{P^0 - \epsilon_1' - \epsilon_2
- \omega_n} + 
 ( 1 \leftrightarrow 2)\right].
\label{eq:Klein-U1++}
\end{eqnarray}
This is the standard time-ordered perturbation theory
(TOPT) one-boson-exchange interaction. Diagrams for the one-pion exchange part of
this interaction between positive-energy nucleon states are 
shown in Fig.~\ref{fig-OPE}. There are additional pieces in the ET
interaction
of Eq.~(\ref{eq:U1ET++}) that arise from $G_C$, i.e., from the
approximate inclusion of higher-order graphs in the crossed-ladder
kernel, which is necessary to obtain the one-body limits.

\section{Construction of three-dimensional nucleon-nucleon interactions}

\label{sec-potentials}

A number of different 3D two-body interactions may be defined by
including or not including retardation effects, and by including, or
not including the contributions of negative-energy states.  These
different 3D interactions are used in our assessment of the significance of
retardations and of negative-energy states in calculations of
electromagnetic observables.  In this section we outline the
development of the various interactions used in this work, in each
case starting from a four-dimensional kernel $K$ that is the sum of
six single-meson exchanges. The mesons are the $\pi(138)$, the
$\sigma(550)$, the $\eta(549)$, the $\rho(769)$, the $\omega(782)$,
and the $\delta(983)$. The quantum numbers and masses of these mesons,
the cutoffs in the (monopole) form factors, as well as the couplings
for all but the $\sigma$ meson, are listed in
Table~\ref{table-mesondata}.  All these parameters except for the
$\sigma $ coupling are taken directly from the Bonn-B fit to the $NN$
phase shifts~\cite{Ma89}. The $\sigma$ coupling is varied so as to
achieve the correct deuteron binding energy for each interaction
considered.

We then use the techniques of the previous section to construct the
following interactions, all of which are to be used in the
three-dimensional ET equation (\ref{eq:ET-Gamma}).
(Note that in calculating the 3D interaction we 
assume that the 4D interaction contains no form factors and any
dependence of vertex factors $V_n$ upon $q^0$ is neglected.  
After 
calculating the three-dimensional interaction we then insert
monopole form factors at all the vertices.)
\begin{enumerate}
\item The ``Retarded ET'' interaction, given by Eq.~(\ref{eq:U1ET}), with
the choices (\ref{eq:equalmasskappa10}) and (\ref{eq:equalmasskappa20})
for $\kappa_1^0$ and $\kappa_2^0$.

\item The instant $NN$ interaction, defined by Eq.~(\ref{eq:U1inst}).

\item The second-order TOPT $NN$ interaction defined by 
Eq.~(\ref{eq:Klein-U1++}).
\end{enumerate} 
The first and second interactions are used in a two-body equation with
the full ET Green's function given by Eq.~(\ref{eq:ET-Gamma}), and
also in an equation in which only the $++$ sector is retained.
However, for the instant interaction, we follow the practice of Devine
and Wallace~\cite{WD94} and switch off by {\it fiat} couplings that
involve flip of both $\rho$-spins, i.e., between the $++$ and $--$
sectors, and between the $+-$ and $-+$ sectors.  A partial
justification of this rule follows from an analysis of the static
limit of our 3D retarded interaction, which shows that all couplings
to the $--$ states vanish as $1/M \rightarrow 0 $, while other
couplings approach the instant form.  Although the static limit for
coupling of $+-$ and $-+$ sectors approaches the standard instant
form, we omit this coupling in order to reproduce the previous results
of Devine and Wallace.

Since, as we have already discussed, the propagator $\langle G_0
\rangle^{-1}$ does not exist in the $+-$ and $-+$ sectors, the TOPT
interaction, $U_{1,{\rm TOPT}}^{++,++}$, is used with only $++$
sectors retained in the equation.

Once a particular interaction is chosen, the integral equation
(\ref{eq:ET-Gamma}) is solved for the bound-state energy.  The method used
involves seeking the energy at which the largest eigenvalue 
of the kernel $U_1 \langle {\cal G}_0 \rangle$
is one. The eigenvalue is calculated using the
Malfliet-Tjon iteration procedure. Details of this method, the rho-spin basis
chosen, and the way the angular integrations are performed may be
found in Ref.~\cite{WD95}. In our calculations we used 40 quadratures
in the radial momentum, and 8 in the integration over the polar angle,
$\theta$. In the case of the two energy-dependent interactions
numerical integration was used to perform the integration over the
azimuthal angle, $\phi$. We found that 12 quadratures were sufficient
to achieve convergence.

In each calculation, the $\sigma$ coupling was adjusted to get the
correct deuteron binding energy, producing the results (accurate to
three significant figures) given in Table~\ref{table-sigmacoupling}.
The value given for the instant calculation with positive-energy
states alone is that found in the original Bonn-B fit. In all other
cases the $\sigma$ coupling must be adjusted to compensate for the
inclusion of retardation, the effects of negative-energy states, etc.
We believe that this adjustment of the scalar 
coupling strength is sufficient to get a reasonable
deuteron wave function. However, one direction for future work is to
refit the meson-exchange parameters in these various different
relativistic $NN$ interaction models to the $NN$ scattering data. From
Table~\ref{table-sigmacoupling} we see that (with the exception
of the TOPT interaction), adjusting the $\sigma$ coupling to
reproduce the observed deuteron binding leads to at most a 5\%
deviation from the Bonn-B value, thereby suggesting that other meson
parameters would change only slightly if a more detailed fit were
performed.

\begin{figure}[h]
\centerline{\BoxedEPSF{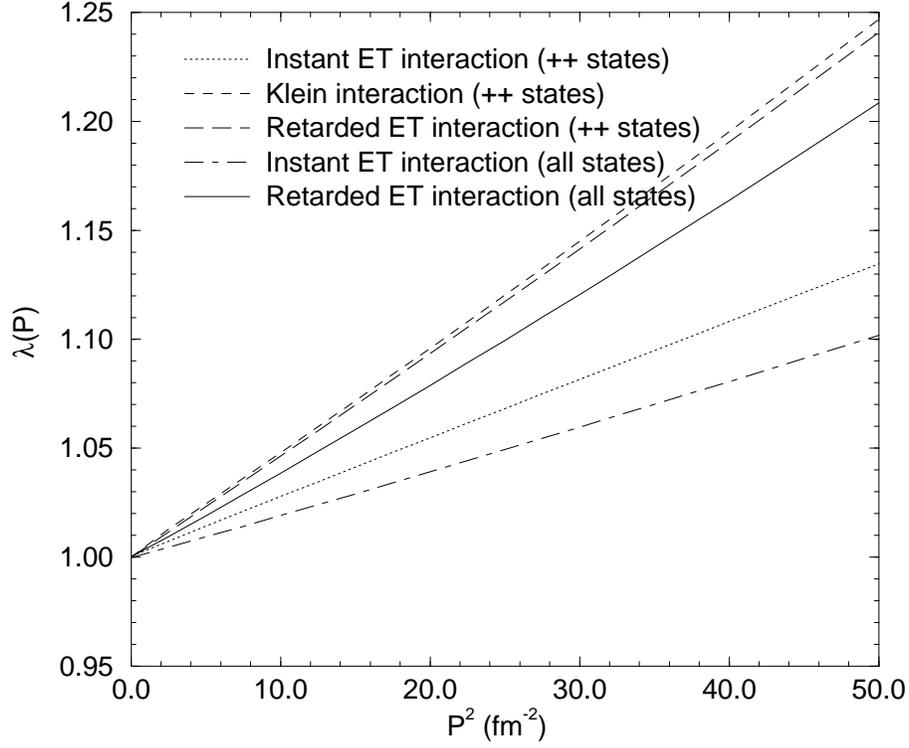 scaled 700}}
\vskip 5 mm
\caption{A plot showing the eigenvalue, $\lambda$, of the homogeneous
integral equation (\protect{\ref{eq:ET}}), as a function of the total
three-momentum of the two-body system, for different choices of the
interaction $U_1$. The dot-dashed (dotted) line is the result when
$U_1$ is chosen to be an instant interaction and all (only $++$)
states are included in the calculation. The solid (long dashed) line
is the result when $U_1$ is chosen to be the retarded interaction
defined by Eq.~(\protect{\ref{eq:U1symm}}) and all (only $++$) states
are included.  Finally, the short-dashed line is the result from the
TOPT interaction, calculated with $++$ states only.}
\label{fig-boost}
\end{figure}

Once the bound-state wave function in the center-of-mass frame has
been determined in this fashion, it is a simple matter to solve the
integral equation (\ref{eq:ET}) in any other frame. As we shall
explain below, we choose to calculate electron-deuteron scattering in
the Breit frame. Hence, we need to calculate the deuteron wave
functions in frames with total four-momentum $(\sqrt{M_D^2 + {\bf
P}^2}, {\bf P})$.  To do this, the interaction 
is recalculated in the new frame using the rules given
above, and then the integral equation is solved with this new
interaction. 
Because the formalism we use for reducing the four-dimensional
integral equation to three dimensions is {\it not} Lorentz invariant
if the potential $U_1$ is truncated at any finite order in the
coupling, we have calculated the eigenvalue
$\lambda({\bf P})$ defined by
\begin{equation} 
U_1({\bf P}) \langle {\cal G}_0(P) \rangle \Gamma_{\rm ET}({\bf P}) 
= \lambda({\bf P}) \Gamma_{\rm ET}({\bf P}),
\end{equation}
for each of the five different interactions defined above.  Since
$\lambda({\bf 0})=1$, by construction, and $P^0 = \sqrt{M_D^2 + {\bf
P}^2}$, is set in accord with Lorentz invariance,  
the deviation of $\lambda$ from one indicates the
violation of Lorentz invariance in the interaction
$U_1$.
The results of this test are shown in Fig.~\ref{fig-boost}.

Three features of these results are worthy of note. Firstly, for
elastic scattering on the deuteron at $Q \approx$5 fm$^{-1}$ in the
Breit frame, the initial and final states have momenta $Q/2 \approx
$2.5 fm$^{-1}$ and $\lambda$ deviates from 1 by only 2\% - 3\%. Thus
for momentum transfers up to of order 1 GeV the violations of Lorentz
invariance are actually quite small.  Secondly, adding meson
retardation to the formalism actually {\it increases} the violation of
Lorentz invariance. This is a little surprising because the
retardation effects, when expanded to order ${\bf P}^2/M^2$, reproduce
the Poincar\'{e} boost operator of Ref.~\cite{Fo95}~\cite{PW96}.  We
note that if the Poincar\'{e} boost operator of order ${\bf P}^2/M^2$
were actually sufficient in this calculation, and effects of higher
order in $1/M$ were truly negligible, then the inclusion of these
retardation effects should remove most of the violation of Lorentz
invariance present in the instant analysis.  This does not happen.
Finally, including negative-energy states in the formalism decreases
the size of the violation of Lorentz invariance.
\section{Construction of a current-conserving electromagnetic interaction}

\label{sec-gi}

In this section, we first review how a conserved electromagnetic
current is developed for the 4D Bethe-Salpeter formalism, and then
show how a corresponding 3D current is obtained from the 4D one.  The
emphasis is on satisfying the appropriate Ward-Takahashi identities.
These reviews establish the procedure that is followed in subsequent
subsections in order to construct the electromagnetic current for the
4D and 3D formalisms involving the propagator $G_C$. The tricky issue
when $G_C$ is present is the construction of the additional terms in
the current to maintain current conservation. Because $G_C$ enters as
a part of two-body interactions associated with crossed graphs, the
current associated with it is a two-body current.  We
construct the extra terms in the current required by the
Ward-Takahashi identity, which is not straightforward because the
variables $\kappa_1^0$ and $\kappa_2^0$ that we choose
depend on the three-momentum of the intermediate state.  As we shall
see below, this means that we cannot use the standard fermion
electromagnetic current, $\gamma_\mu$, because such a choice would
violate current conservation.  We show that the Ward-Takahashi identity
may be recovered by incorporating an additional term in the current.
This yields the ET current referred to as ${\cal G}^\gamma_{0,\mu}$ in
the Introduction. We then develop the corresponding 3D current for the
formalism involving $G_C$. Because the ET current has parts
attributable to two-body effects, we also develop a simpler current
that is appropriate for use with instant interactions, ${\cal
G}^{\gamma}_{{\rm inst} \mu}$.  This last-mentioned current is the one
used in the calculations of this paper, which are based upon the
impulse approximation.  Although we do not use the full formalism
developed in this section for calculations in this paper, the results
presented here are pertinent to future calculations in which two-body
currents will be included.

\subsection{Review of Ward-Takahashi identities in 4D Bethe-Salpeter formalism}

We begin our discussion by reviewing the WTI for the usual
Bethe-Salpeter Green's function.  Consider the two-body Green's
function $G$ that is the solution of the Bethe-Salpeter equation
\begin{equation}
G=G_0 + G_0 K G,
\label{eq:GBSE}
\end{equation}
where $K$ is any BSE kernel. Define the two-body Green's function for
the interaction of two free particles with a photon of fixed momentum
$Q$ via:
\begin{eqnarray}
G_{0,\mu}^\gamma(p_1,p_2,Q) &\equiv& 
G_0(p_1,p_2+Q) [-i j^{(2)}_\mu(Q^2)
d_1^{-1}(p_1)] G_0(p_1,p_2) + ( 1 \leftrightarrow 2) \nonumber \\
&=& G_{0, \mu}^{\gamma (2)}(p_1,p_2,Q) + G_{0,\mu}^{\gamma
(1)}(p_1,p_2,Q).
\label{eq:tbgamma}
\end{eqnarray}
Here, and throughout the rest of the paper, the notation $(1
\leftrightarrow 2)$ indicates that the momenta of the two
particles must be swapped, {\it and} the labels exchanged. Therefore,
the $(1 \leftrightarrow 2)$ pieces of any expression represent the
photon coupling to whichever particle it did not couple in the first
part of the expression. An explicit example of this rule is the
particle one coupling term of Eq.~(\ref{eq:tbgamma}), which is
$G_{0,\mu}^{\gamma(1)}(p_1,p_2,Q) = G_0(p_1+Q,p_2) [-i d_2^{-1}(p_2)
j_\mu^{(1)}(Q^2)] G_0(p_1,p_2)$.

The free Green's function 
$G_{0,\mu}^\gamma$ obeys a Ward-Takahashi identity.
Now let $G_\mu^\gamma$ be the Green's function for the interaction of
one photon with the interacting two-particle system.  Note that in the
two-nucleon system the charges $q_i$ will include isospin
operators, so care must be exercised in ordering charges and
isospin-dependent interactions. We may write the following equation
for a current-conserving $G_\mu^\gamma$, by allowing the photon to be
inserted anywhere on the right-hand side of Eq.~(\ref{eq:GBSE}), and
then rearranging the result:
\begin{eqnarray}
&&G^\gamma_\mu(k_1',k_2';k_1,k_2;Q)=\int \frac{d^4p_1 d^4p_2 }{(2
\pi)^8} G(k_1',k_2';p_1,p_2+Q) [-i j^{(2)}_\mu(Q^2) d_1^{-1}(p_1)]
G(p_1,p_2;k_1,k_2)\nonumber\\ 
&& \quad + (1 \leftrightarrow 2) + \int
\frac{d^4p_1' d^4p_2' d^4p_1 d^4p_2}{(2 \pi)^{16}}
G(k_1',k_2';p_1',p_2') K^\gamma_\mu(p_1',p_2';p_1,p_2;Q)
G(p_1,p_2;k_1,k_2).
\label{eq:Ggameq2}
\end{eqnarray}
Here $K^\gamma_\mu$ is found by coupling the photon to every internal
charged line in the kernel $K$. (Note that there is an overall delta
function $\delta^{(4)} (k_1' + k_2' - k_1 - k_2 - Q)$ on both sides of
this equation.) Using the WTI for $G_{0,\mu}^\gamma$,
Eq.~(\ref{eq:Ggameq2}), and the explicit form of Eq.~(\ref{eq:GBSE})
we find a WTI for $G_\mu^\gamma$,
\begin{equation}
Q^\mu G^\gamma_\mu(k_1',k_2';k_1,k_2;Q)=q_2 G(k_1',k_2'-Q;k_1,k_2) -
G(k_1',k_2';k_1,k_2+Q) q_2 + (1 \leftrightarrow 2),
\end{equation}
provided that
\begin{equation}
q_2 K(p_1',p_2'-Q;p_1,p_2) - K(p_1',p_2';p_1,p_2+Q) q_2 + (1
\leftrightarrow 2)=Q^\mu K^\gamma_\mu (p_1',p_2';p_1,p_2;Q),
\label{eq:K-WTI}
\end{equation}
which is the WTI for the interaction current.  (Similar identities are
used in the construction of Ward-Takahashi identities for the
Gross---or spectator---formalism in Refs.~\cite{GR87,KB97A}.) The
result (\ref{eq:K-WTI}) is completely general, and will always hold if
the two-body current $K^\gamma_\mu$ is constructed by coupling
the photon to every charged line in the kernel $K$.

Using the usual decomposition of the two-body Green's function into a
pole and a regular part gives the amplitude for interaction of the
bound-state with a photon of momentum $Q$~\cite{Ma55}.  Expressing the
result in terms of total and relative four-momenta yields:
\begin{eqnarray}
&&{\cal A}_\mu(P,Q)=\int \frac{d^4p}{(2 \pi)^4} 
\bar{\Gamma}(p-Q/2;P') G^{\gamma (2)}_{0,\mu} (p;P;Q) 
\Gamma(p;P). + (1 \leftrightarrow 2) \nonumber\\
&& \qquad + \int \frac{d^4p' \, d^4p}{(2 \pi)^8} 
\bar{\Gamma}(p';P') G_0(p';P') K_\mu^\gamma (p',P';p,P;Q)
G_0(p;P) \Gamma(p;P),
\label{eq:gi4damp}
\end{eqnarray}
where $P' = P+Q$ and $\Gamma(p;P)$ is the two-body vertex function corresponding to
the bound-state at four-momentum $P^2=M^2$.  From the WTIs for
${G_{0}^\gamma}_\mu$ and $K_\mu^\gamma$, and the bound-state BSE,
$\Gamma = K G_0 \Gamma$, it is then straightforward to show that
\begin{equation}
Q^\mu {\cal A}_\mu(P,Q)=0,
\label{eq:BSWTI}
\end{equation}
as required for current conservation.

\subsection{Gauge invariance in a 3D reduction of the Bethe-Salpeter
formalism}

\label{sec-3Dcurrents}

In the context of a reduction to three dimensions, the question is how
to maintain gauge invariance when the reduction is made. In
Section~\ref{intro-summary} we outlined the 3D reduction method for
the bound-state vertex arising from the solution of the Bethe-Salpeter
equation. This gives
\begin{equation}
\Gamma_1(P) = K_1(P) {\bf \langle} G_0(P) {\bf \rangle} \Gamma_1 (P), 
\label{eq:3Deqn}
\end{equation}
 where the dependence of quantities on the total four momentum, $P$, is
indicated.  The 4D vertex function, $\Gamma$, and the corresponding 3D
one, $\Gamma_1$, are related as follows,
\begin{equation}
G_0(P) \Gamma(P) = {\cal G}(P) \Gamma_1(P).
\label{eq:GcalGamma1}
\end{equation}
Moreover, the interaction $K_1$ is defined so as to
obey, 
\begin{equation}
\langle G_0(P) \rangle K_1(P) \langle G_0 (P) \rangle = \langle G_0(P) K(P)
{\cal G}(P) \rangle.
\label{eq:calGeq}
\end{equation}  

Inserting Eq.~(\ref{eq:GcalGamma1}) into Eq.~(\ref{eq:gi4damp}), we
find an entirely equivalent expression for the current,
\begin{eqnarray}
&& {\cal A}_\mu(P,Q)=\int \frac{d^4k' d^4p' d^4p \, d^4k}{(2 \pi)^{16}} 
\bar{\Gamma}_1({\bf k}';P') 
{\cal G}(k',p';P') \nonumber\\ 
&& \left[-i j_\mu^{(1)}(Q^2)
d_2^{-1}(\nu_2 P - p) \delta^{(4)}(p'-p-Q/2) -i j_\mu^{(2)}(Q^2)
d_1^{-1}(\nu_1 P + p) \delta^{(4)}(p'-p+Q/2)) \right.\nonumber\\ &&
\qquad \qquad \qquad \qquad \qquad \qquad \left. + K^\gamma_\mu (P',p';P,p;Q) 
\right] {\cal G}(p,k;P) \Gamma_1({\bf k};P).
\label{eq:3damplfull}
\end{eqnarray}
Because the vertex functions do not depend upon the time-components of
relative momenta, integrations over $k^0$ and $k^{0'}$ reduce this
expression to a 3D one, which we abbreviate as 
\begin{equation}
{\cal A}_\mu=\bar{\Gamma}_1(P') \langle {\cal G}(P') 
\left[J_\mu + K_\mu^\gamma \right] {\cal G}(P) \rangle \Gamma_1(P).
\label{eq:Amu}
\end{equation}

Now, given a result for $\Gamma_1$ obtained by systematic expansion
of $K_1$, the amplitude ${\cal A}_\mu$ also can be expanded
systematically in a way that maintains current conservation.  First,
note that we solve for $K_1$ in accord with Eq.~(\ref{eq:calGeq}).
This leads to an infinite series for $K_1$.  If the condition
(\ref{eq:calGeq}) is imposed order-by-order in the expansion in
$K-K_1$, the condition defines $K_1$ to the same order.  Truncation of
the kernel is necessary for a practical analysis and we must ask if a
corresponding 3D approximation for the current matrix element
(\ref{eq:Amu}) exists that maintains the Ward-Takahashi identities of
the theory.  {\it It turns out that the current matrix element
(\ref{eq:Amu}) is conserved if ${\cal G} (J_\mu + K^\gamma_\mu) {\cal
G}$ on the right-hand side of Eq.~(\ref{eq:Amu}) is expanded to a
given order in the coupling constant and the kernel $K_1$ used to
define $\Gamma_1$ is obtained from Eq.~(\ref{eq:calGeq}) by truncation
at the same order in the coupling constant.}

To effect this the right-hand side of Eq.~(\ref{eq:Amu}) is split into
two pieces, one due to the one-body current $J_\mu$, and one due to
the two-body current $K^\gamma_\mu$. Suppose now that $K_1$ has been
truncated at lowest order, i.e., $K_1=K_1^{(2)}$, and that
$K^\gamma_\mu = K^{\gamma (2)}_\mu$.  Then, in the $J_\mu$ piece, we
expand the $\cal G$s and retain terms up to the same order in
$K^{(2)}-K_1^{(2)}$.  A piece from the two-body current must be added
to this.  In that piece we stop this expansion of $\cal G$ at zeroth
order in $K^{(2)}-K_1^{(2)}$, i.e., write ${\cal G}=G_0$.  Thus, we
define our second-order approximation to ${\cal A}_\mu$, ${\cal
A}_\mu^{(2)}$, by
\begin{eqnarray}
{\cal A}^{(2)}_\mu&=&\bar{\Gamma}_1(P') \langle G^\gamma_{0 \mu}
\rangle \Gamma_1(P) \nonumber \\ &+& \bar{\Gamma}_1(P') \langle G_0(P') (K^{(2)}(P')-K^{(2)}_1(P'))
G^\gamma_{0 \mu} \rangle \Gamma_1(P)
\nonumber\\
&+& \bar{\Gamma}_1(P') \langle G^{\gamma}_{0 \mu} (K^{(2)}(P)-K^{(2)}_1(P)) G_0(P) \rangle
 \Gamma_1(P)
\nonumber\\ 
&+& \bar{\Gamma}_1(P') \langle G_0(P') K^{\gamma (2)}_\mu G_0(P) \rangle
\Gamma_1(P).
\label{eq:A2mu}
\end{eqnarray}
To show that ${\cal A}^{(2)}_\mu$ is gauge invariant, it must be
contracted with the four-vector $Q$. The WTIs for
${G_0}_\mu^\gamma$, $K_\mu^\gamma$, together with the bound-state
equation (\ref{eq:3Deqn}), can be used to show that
\begin{eqnarray}
Q^\mu {\cal A}^{(2)}_\mu=
-\bar{\Gamma}_1(P') \langle G_0(P') (K^{(2)}(P')-K^{(2)}_1(P')) G_0(P') \rangle q_1
\Gamma_1(P)\nonumber\\
+ \bar{\Gamma}_1(P') q_1 \langle G_0(P) (K^{(2)}(P)-K^{(2)}_1(P)) G_0(P) \rangle
\Gamma_1(P)+ (1 \leftrightarrow 2).
\end{eqnarray}
Thus, at second-order in the coupling constant, with
Eq.~(\ref{eq:calGeq}) expanded at second order defining $K_1^{(2)}$,
the corresponding amplitude for electromagnetic interactions of the
bound state, as defined by Eq.~(\ref{eq:A2mu}), obeys
\begin{equation}
Q^\mu {\cal A}^{(2)}_\mu=0.
\label{eq:A1WTI}
\end{equation}

It is straightforward to check that the same result holds if
Eq.~(\ref{eq:calGeq}) for $K_1$ is truncated at fourth order, while
the one-body and two-body current pieces of Eq.~(\ref{eq:3damplfull})
similarly are expanded to fourth order.  This defines a vertex
function ${\cal A}_\mu^{(4)}$ which obeys $Q^\mu {\cal
A}^{(4)}_\mu=0$. Thus, truncating the kernel $K_1$ defined by
Eq.~(\ref{eq:calGeq}) and the electromagnetic vertex defined by
Eq.~(\ref{eq:Amu}) at consistent order in the coupling constant yields
a current-conserving electromagnetic matrix element.  A covariant
extension of the formalism presented here is given in
Ref.~\cite{PW97}.

In fact, ${\cal A}_\mu^{(2)}$ includes contributions from diagrams
where the photon couples to particles one and two while exchanged
quanta are ``in-flight''. These contributions are of two kinds.
Firstly, if the four-dimensional kernel $K$ is dependent on the total
momentum, or if it involves the exchange of charged particles, then
gauge invariance requires the presence of terms representing the
coupling of the photon to internal lines in $K$.  Secondly, even if
gauge invariance does not require the presence of such terms in the
four-dimensional formalism, terms arise in the three-dimensional
formalism where the photon couples to particles one and two while an
exchanged particle is ``in-flight''. These must be included if our
approach is to contain a WTI. (See Fig.~\ref{fig-inflight} for a
diagrammatic interpretation of one such term.)

A special case of the above results occurs when retardation effects are
omitted, i.e., the kernel $K_1=K_{\rm
inst}$, is chosen, and the bound-state equation (\ref{eq:Salpeter}) is
solved to get the vertex function $\Gamma_1=\Gamma_{\rm inst}$. Then a
gauge-invariant current is obtained by implementing the instant
approximation in the expression (\ref{eq:A2mu}) in a way consistent
with that in which it was used in obtaining the Salpeter equation
(\ref{eq:Salpeter}). Taking Eq.~(\ref{eq:A2mu}) and imitating
the derivation of the Salpeter equation by replacing $K$ by
$K_{\rm inst}$ leads to
\begin{equation}
{\cal A}_{{\rm inst},\mu}(P,Q)=\bar{\Gamma}_{\rm inst}(P') \langle
G_{0 \mu}^\gamma \rangle \Gamma_{\rm inst}(P) + \bar{\Gamma}_{\rm
inst}(P') \langle G_0(P') \rangle {K^\gamma_{\rm inst}}_\mu 
\langle G_0(P) \rangle \Gamma_{\rm inst}(P),
\label{eq:instantme}
\end{equation}
where we have also replaced the meson-exchange current kernel
$K^\gamma_\mu$ by the instant approximation thereto. By definition
this instant meson-exchange current kernel obeys the WTI
(\ref{eq:K-WTI}), but with the instant kernel $K_{\rm inst}$
appearing on the right-hand side.  To show that
Eq.~(\ref{eq:instantme}) defines a gauge-invariant matrix element we
contract with the four vector $Q^\mu$ and use the Ward-Takahashi
identities for $J_\mu$ and $K^\gamma_{{\rm inst} \mu}$. This gives:
\begin{eqnarray}
&& Q^\mu {\cal A}_{{\rm inst},\mu}(P,Q)=\bar{\Gamma}_{\rm inst}(P') q_2
\langle G_0(P) - G_0(P') \rangle \Gamma_{\rm inst}(P) 
\nonumber\\
&& \qquad \qquad + \bar{\Gamma}_{\rm inst}(P') \langle G_0(P')
\left(q_2 K_{\rm inst}(P) - K_{\rm inst}(P')q_2 \right) G_0(P) \rangle
\Gamma_{\rm inst}(P) + (1 \leftrightarrow 2).
\end{eqnarray}
Using the bound-state equation, $\Gamma_{\rm inst} = K_{\rm inst}
\langle G_0 \rangle \Gamma _{\rm inst}$, both at total momentum $P$ and at 
total momentum $P'$ then gives the desired WTI:
\begin{equation}
Q^\mu {\cal A}_{{\rm inst},\mu}(P,Q)=0.
\label{eq:instWTI}
\end{equation}
Note that if the four-dimensional interaction $K$ does not depend on 
the total momentum $P$ then the simple current
\begin{equation}
{\cal A}_{{\rm inst},\mu}(P,Q)=\bar{\Gamma}_{\rm inst}(P')
\langle G^\gamma_{0,\mu}(P;Q) \rangle \Gamma_{\rm inst}(P),
\label{eq:simpleinstant}
\end{equation}
is gauge invariant.

\subsection{Gauge invariance in the 4D formalism with $G_C$}

In this section we construct a conserved current for
the 4D equation
\begin{equation}
\Gamma=U {\cal G}_0 \Gamma.
\label{eq:4DET2}
\end{equation}
In order to do this we rewrite the equation as two coupled
equations, the first of which involves
the Bethe-Salpeter kernel $K$, 
\begin{equation}
\Gamma=K G_0 \Gamma,
\label{eq:4DETalt}
\end{equation}
and the second of which defines the reduced interaction $U$, 
\begin{equation}
K = U + U G_C K.  
\end{equation}
From Eq.~(\ref{eq:gi4damp}) we see that the gauge-invariant
current for the photon coupling to the bound state will contain a
piece corresponding to the coupling of photon lines in all possible
ways inside $K$. This piece of the current is composed of two
parts, as follows,
\begin{equation}
K^\gamma_{\mu} = (1 + K G_C )U^\gamma_\mu (1 + G_C K) + 
K G_{C,\mu}^\gamma K.
\end{equation} 
The Ward-Takahashi identities that are required to be satisfied
involve one for $U^\gamma_\mu$ that takes the same form as
Eq.~(\ref{eq:K-WTI}), with $K$ replaced by $U$.  
\begin{equation}
q_2 U(p_1',p_2'-Q;p_1,p_2) - U(p_1',p_2';p_1,p_2+Q) q_2 + (1
\leftrightarrow 2)=Q^\mu U^\gamma_\mu (p_1',p_2';p_1,p_2;Q).
\label{eq:U-WTI}
\end{equation}
The other involves the Green's function $G^\gamma_{C,\mu}$,
representing a photon coupling to the propagator $G_C$. This piece
must be constructed in accordance with the Ward-Takahashi identity:
\begin{equation}
Q^\mu G^\gamma_{C,\mu}(p_1,p_2,Q)=q_2 (G_C(p_1,p_2) - G_C(p_1,p_2+Q))
+ (1 \leftrightarrow 2).
\label{eq:GCWTI}
\end{equation}
When both of the WTI's (\ref{eq:U-WTI}) and (\ref{eq:GCWTI}) hold,
then the electromagnetic kernel $K^\gamma_\mu$ will satisfy the WTI
(\ref{eq:K-WTI}). In this case the current appropriate to a photon
insertion in the free propagator is
\begin{equation}
{\cal G}_{0,\mu}^\gamma \equiv G_{0,\mu}^\gamma +
G_{C,\mu}^\gamma,
\end{equation}
and so, if $\Gamma$ obeys Eq.~(\ref{eq:4DET2}), the amplitude
\begin{eqnarray}
&&{\cal A}_\mu(P,Q)=\int \frac{d^4p}{(2 \pi)^4} 
\bar{\Gamma}(p-Q/2;P') {\cal G}^{\gamma (2)}_{0,\mu} (p;P;Q) 
\Gamma(p;P) + (1 \leftrightarrow 2) \nonumber\\
&& \qquad + \int \frac{d^4p' \, d^4p}{(2 \pi)^8} 
\bar{\Gamma}(p';P') {\cal G}_0(p';P') U_\mu^\gamma (p',P';p,P;Q)
{\cal G}_0(p;P) \Gamma(p;P),
\label{eq:gi4damp2}
\end{eqnarray}
will satisfy the current conservation condition (\ref{eq:BSWTI}). 

In order to implement this formalism, we must construct a current
${G^\gamma_C}_\mu$ that obeys Eq.~(\ref{eq:GCWTI}).  Rather than
immediately tackling the full $G_C$ defined by Eq.~(\ref{eq:G0plusGC})
we first perform this task for one term by considering the simple case
in which the propagator is $i d_1 d_2^c$ and the photon couples only
to particle two. This is non-trivial because the choice of variable
$\kappa_2^0$ in $d_2^c$ in general depends on the three-momenta of
both particles. For instance, if $\kappa_2^0$ is chosen as in
Eq.~(\ref{eq:equalmasskappa20}) and the particles are of equal mass
then the $\kappa_2^0$ which appears in the propagator $G_C(p_1,p_2 +
Q)$ is (for equal-mass particles)
\begin{equation}
{\kappa_2^0}'=\frac{1}{2} (P_0 + Q_0 - \epsilon_1 + \epsilon_2'),
\label{eq:kappa20p}
\end{equation}
where $P \equiv p_1 + p_2$, and $\epsilon_2'=\epsilon_2({\bf p}_2 +
{\bf Q})$.  This is to be compared to the case where $\kappa_2^0$ is
defined via Eq.~(\ref{eq:nu2P0}). In that case the $\kappa_2^0$
appearing in $G_C(p_1,p_2+Q)$ is, for equal mass particles,  
\begin{equation}
\kappa_2^{0 \prime}={1 \over 2} (P_0 + Q_0),
\label{eq:kappa20pold}
\end{equation}
and there is thus no dependence on the intermediate-state
three-momenta of particles one and two.  As we shall see below, the
appearance of momenta in the choice (\ref{eq:kappa20p}) for
$\kappa_2^0$ leads to complications in the construction of a current
obeying (\ref{eq:GCWTI}).

To construct such a current, we first postulate the form
\begin{equation}
d^{\gamma (2)}_{c,\mu}(p_2;Q) \equiv {d_2^{\tilde c}}(p_2+Q) j^{(2)}_{c,\mu}
d_2^c(p_2),
\end{equation}
where we write $d_2^{\tilde c}$ in order to indicate that the constraint
on $\kappa_2^0$ is different once the photon strikes
particle two. The WTI for this current is then:
\begin{equation}
Q^\mu j^{(2)}_{c,\mu}=q_2 (d_2^{\tilde{c}}(p_2 + Q)^{-1} - d_2^c(p_2)^{-1}).
\label{eq:d2cWTI}
\end{equation}
Constructing the right-hand side of this equation, for spin-half particles
we obtain
\begin{equation}
q_2 \left[2(\not \! \kappa_2' - \not \! \kappa_2) - \not \! Q\right]
\end{equation}
Using the form for $\kappa_2$, Eq.~(\ref{eq:kappa2}), and three-momentum
conservation, ${\bf p}_2'={\bf p}_2 + {\bf Q}$, we see that
\begin{equation}
d_2^{\tilde{c}}(p_2 + Q)^{-1} - d_2^c(p_2)^{-1}=\left[2({\kappa_2^0}'
- \kappa_2^0) - 2 Q^0 \right] \gamma_2^0 + \not \! Q.
\end{equation}
Note that if we make the choice (\ref{eq:nu2P0}) then, $\kappa_2^{0
\prime}$ is given by $\kappa_2^{0 \prime}=\nu_2 (P^0 + Q^0)$ and
$\kappa_2^0=\nu_2 P^0$, so
\begin{equation}
d_2^{\tilde{c}}(p_2 + Q)^{-1} - d_2^c(p_2)^{-1}=\not \! Q +2(\nu_2 - 1)
Q^0 \gamma_2^0.
\end{equation}
This means that, in the Breit frame where $Q^0= 0$, the 
usual current $j^{(2)}_\mu= q_2 {\gamma_2}_\mu$ satisfies
the Ward-Takahashi identity (\ref{eq:d2cWTI}).

However, when we use the formulae (\ref{eq:kappa20p}) and
(\ref{eq:equalmasskappa20}) in the case of equal-mass particles, note that 
\begin{equation}
2({\kappa_2^0}' - \kappa_2^0)=Q^0 + \frac {{\bf Q} 
\cdot ({\bf p}_2' + {\bf p}_2)}{\epsilon_2 + \epsilon_2'},
\end{equation}
thus implying that
\begin{equation}
d_2^{\tilde{c}}(p_2 + Q)^{-1} - d_2^c(p_2)^{-1}=\not \! Q - Q \cdot 
\frac{\hat{p}_2' + \hat{p}_2}{\epsilon_2' + \epsilon_2} {\gamma_2}_0,
\end{equation}
where $\hat{p}_2'=(\epsilon_2',{\bf p}_2')$, 
$\hat{p}_2=(\epsilon_2,{\bf p}_2)$. Consequently, the WTI
(\ref{eq:d2cWTI}) requires that we define the current
\begin{equation}
j^{(2)}_{c,\mu}= q_2 \gamma_\mu - \tilde{j}^{(2)}_\mu,
\end{equation}
where
\begin{equation}
\tilde{j}^{(2)}_\mu=q_2 \frac{\hat{p}_{2 \mu}' + \hat{p}_{2
\mu}}{\epsilon_2' + \epsilon_2} {\gamma_{2}}_0,
\label{eq:tildej2}
\end{equation}
is the additional piece of the current that compensates for the
dependence of $\kappa_2^0$ on momentum.  Note that
$j^{(2)}_{c,0} = 0$ with this choice of $j^{(2)}_{c,\mu}$. Note also the
identity
\begin{equation}
Q^\mu \tilde{j}^{(2)}_{\mu}=2 ({\kappa_1^0}' - {\kappa_1^0}) q_2 {\gamma_2}_0.
\end{equation}

Next we consider a second piece of $G_C$, namely 
the propagator $i d_1^c d_2$, still
in the situation where the photon couples only to particle two. Denote
the free $N N \rightarrow N N \gamma$ Green's function corresponding
to this situation by $d^{\tilde{c}}_\mu$.  The
Ward-Takahashi identity which this Green's function should obey
is, 
\begin{equation}
Q^\mu d^{\tilde c}_\mu(p_1,p_2;Q)=i q_2 [d_1^c(p_1) d_2(p_2) -
d_1^{\tilde c}(p_1) d_2(p_2+Q)].
\label{eq:specWTI}
\end{equation}
Note that here the propagator $d_1^{\tilde{c}}(p_1) \neq d_1^c(p_1)$,
since the change in the momentum of particle two after the impact of
the photon leads to a change in the choice of $\kappa_1^0$ that
appears in $d_1^{\tilde c}$---in spite of particle one apparently
being a spectator in this interaction. 
 Again, the simplest choice
\begin{equation}
d^{\tilde c}_\mu(p_1,p_2;Q)=i d_1^c(p_1) d_2(p_2 + Q) q_2 \gamma_{2
\mu} d_2(p_2)
\end{equation}
does not obey the Ward-Takahashi identity. However, if we construct
$d^{\tilde c}_\mu$ according to 
\begin{eqnarray}
d^{\tilde c}_\mu(p_1,p_2;Q)&=&\frac{i}{2}\left[(d_1^c(p_1) +
d_1^{\tilde c}(p_1)) d_2(p_2 + Q) q_2 \gamma_{2 \mu} d_2(p_2) \right.
\nonumber\\ 
&+& \left. d_1^{\tilde c}(p_1) \tilde{j}_{2 \mu} \gamma_2^0
\gamma_1^0 d_1^c(p_1) (d_2(p_2 + Q) + d_2(p_2)) \right]
\label{eq:extrabit}
\end{eqnarray}
then it does obey the WTI (\ref{eq:specWTI}). This can be
shown as follows. Contracting the right-hand side of
Eq.~(\ref{eq:extrabit}) with $Q^\mu$ gives
\begin{eqnarray*}
&&\frac{i}{2}\left[(d_1^c(p_1) + d_1^{\tilde c}(p_1)) q_2 (d_2(p_2) -
d_2(p_2 + Q))\right.\\ 
&& \qquad \qquad \left.+ d_1^{\tilde c}(p_1) q_2 2({\kappa_1^0}' -
\kappa_1^0) \gamma_1^0 d_1^c(p_1) (d_2(p_2 + Q) + d_2(p_2))\right].
\end{eqnarray*}
But, 
\begin{equation}
2({\kappa_1^0}' - \kappa_1^0) {\gamma_1}_0={d_1^{\tilde c}}(p_1)^{-1}
- {d_1^c}(p_1)^{-1},
\end{equation}
and so, the Green's function defined by Eq.~(\ref{eq:extrabit}) does
indeed satisfy the WTI (\ref{eq:specWTI}). 

It is now possible to combine the results for $j^{(2)}_{c,\mu}$ and $d^{\tilde
c}_\mu$ in order to construct the 4D current ${\cal
G}_{0,\mu}^\gamma = {G_0^\gamma}_\mu +
{G_C^\gamma}_\mu$ corresponding to the free Green's function
(\ref{eq:G0plusGC}).  In this case the relevant WTI (again, only
writing explicitly the parts where the photon couples to particle two)
is:
\begin{eqnarray}
&& Q^\mu {\cal G}^\gamma_{0,\mu}(p_1,p_2;Q) =
\frac{i}{2}q_2(d_1(p_1) + d_1^c(p_1)) (d_2(p_2) + d_2^c(p_2)) \nonumber\\
&& \qquad \qquad \qquad - \frac{i}{2}q_2(d_1(p_1) + d_1^{\tilde c}(p_1))
(d_2(p_2+Q) + d_2^{\tilde c} (p_2+Q)) + (1 \leftrightarrow 2).
\label{eq:fullWTI}
\end{eqnarray}
Using the forms already constructed for $j^c_{2 \mu}$ and
$d_\mu^{\tilde c}$ we discover that the 4D current in the
formalism involving $G_C$ is:  
\begin{eqnarray}
&&{\cal G}^\gamma_{0,\mu}(p_1,p_2;Q)
=\nonumber\\ && \quad \frac{i}{4} \left[(2 d_1(p_1)
+ d_1^c(p_1) + d_1^{\tilde c}(p_1)) (d_2(p_2 + Q) q_2\gamma_{2\mu} d_2(p_2)
+ d_2^{\tilde c}(p_2+Q) j^{(2)}_{c,\mu} d_2^c(p_2)) \right.\nonumber\\ && 
\quad \quad + \left. d_1^{\tilde c}(p_1) \tilde{j}_{2 \mu} \gamma_2^0
\gamma_1^0 d_1(p_1) (d_2(p_2) + d_2(p_2 + Q) + d_2^c(p_2) +
d_2^{\tilde c}(p_2 + Q)) \right] + (1 \leftrightarrow 2),
\label{eq:giprop}
\end{eqnarray}
and it obeys Eq.~(\ref{eq:fullWTI}).

\subsection{ Reduction to 3D and the ET current} 

Having constructed a 4D current for the formalism involving $G_C$ that
obeys the required Ward-Takahashi identity, we can apply the reduction
formalism of Section~\ref{sec-3Dcurrents} to obtain the currents
corresponding to the 3D reduction of this 4D theory.  To this end, we
first note that the 4D vertex functions are related to 3D ones,
$\Gamma _{\rm ET}$, by Eqs.~(\ref{eq:OmegaRGamma}) and
(\ref{eq:GammaOmegaL}).  Using these relations to rewrite the 4D
current (\ref{eq:AmuET}) in terms of the 3D vertex functions shows that
\begin{eqnarray}
{\cal A}_{{\rm ET},\mu}&=&\bar{\Gamma}_{1,{\rm ET}}(P') 
\langle \Omega^L (P') {\cal G}^\gamma_{0,\mu} 
 \Omega^R (P) \rangle \Gamma_{1,{\rm ET}}(P) \nonumber\\
\nonumber\\ 
&+& \bar{\Gamma}_{1,{\rm ET}}(P') \langle {\cal G}(P') U^{\gamma }_\mu 
{\cal G}(P) \rangle \Gamma_{1,{\rm ET}}(P).
\label{eq:general3DETcurrent}
\end{eqnarray}
Here, $U^\gamma_\mu$ denotes the interaction current obtained by
coupling the photon in all possible ways within the reduced
interaction $U$.  In the same manner as discussed for the 3D version
of the Bethe-Salpeter current, it is possible to maintain gauge
invariance if the 3D interaction $U_1$ is truncated at some definite
order in the coupling constant, while the expansions for $\Omega^L
{\cal G}_{0,\mu}^\gamma \Omega^R$ and ${\cal G} U^\gamma_\mu {\cal G}$
similarly are truncated at the same order in the coupling constant.

Because we truncate our effective interaction at second order,
the corresponding conserved current in 3D form is obtained by
use of $\Omega^{R(2)} = 1 + 
 [U^{(2)} - U_1 ^{(2)}]{\cal G}_0 $ and a corresponding expansion
for $\Omega^{L(2)}$ in Eq.~(\ref{eq:general3DETcurrent}).  The result for
the 3D current, 
\begin{eqnarray} 
{\cal A}^{(2)}_\mu&=&\bar{\Gamma}_{1,{\rm ET}}(P') \langle {\cal
G}^\gamma_{0,\mu} \rangle \Gamma_{1,{\rm ET}}(P) \nonumber\\
&+& \bar{\Gamma}_{1,{\rm ET}}(P') \langle
{\cal G}_0(P') (U^{(2)}(P')-U_1^{(2)}(P')) {\cal G}^\gamma_{0,\mu} 
\rangle \Gamma_{1,{\rm ET}}(P)
\nonumber\\
&+& \bar{\Gamma}_{1,{\rm ET}}(P') \langle {\cal G}^{\gamma}_{0,\mu} 
(U^{(2)}(P)-U_1^{(2)}(P)) {\cal G}_0(P) \rangle \Gamma_{1,{\rm ET}}(P)
\nonumber\\ 
&+& \bar{\Gamma}_{1,{\rm ET}}(P') \langle {\cal G}_0(P') U^{\gamma (2)}_\mu 
{\cal G}_0(P) \rangle \Gamma_{1,{\rm ET}}(P),
\label{eq:general3DETcurrent(2)}
\end{eqnarray}
obeys the Ward-Takahashi identity (\ref{eq:A1WTI}).

Note that $U^{(2)} = K^{(2)}$ and $U_\mu^{\gamma (2)} = K_\mu^{\gamma
(2)}$ are correct at second order in expansions in the coupling
constant. Consequently if we are considering one-boson exchange
interactions the only contributions to $K_\mu^{\gamma (2)}$ that are
necessary for gauge invariance come from the photon coupling to
isovector exchange particles and from contact terms due to derivative
couplings of the mesons to the nucleon. Since these give rise to 
isovector structures their contribution to electromagnetic scattering
off the deuteron is, in fact, zero.

\subsection{Impulse-approximation current based on the instant
approximation to ET formalism}

Just as in the case of the Bethe-Salpeter equation, if the instant
approximation is used to obtain from Eq.~(\ref{eq:4DET}) the
three-dimensional equation that defines the vertex function, then a
corresponding gauge-invariant impulse current can be constructed by
replacing $U$ by $U_{\rm inst}$ everywhere in the expression
(\ref{eq:general3DETcurrent(2)}). Assuming that the interaction in the
four-dimensional equation does not depend on the total momentum, this
leads to a particularly simple conserved current:
\begin{equation}
{\cal A}_{{\rm inst},\mu}=\bar{\Gamma}_{\rm inst} 
\langle {\cal G}^\gamma_{0,\mu} \rangle
\Gamma_{\rm inst}.
\end{equation}

At this point it appears that we must integrate the rather complicated
formula (\ref{eq:giprop}) over the zeroth component of relative
momentum in order to calculate $\langle {\cal G}^\gamma_{0,\mu} 
\rangle$. However, the result (\ref{eq:giprop}) for
${\cal G}^\gamma_{0,\mu}$ was constructed in order to
obey Ward-Takahashi identities in the full four-dimensional theory. It is
not, in fact, necessary to use the full result if we are only
concerned with maintaining WTIs at the three-dimensional
level in the instant approximation.  As remarked
following Eq.~(\ref{eq:aveG0GC}),  the simpler propagator
$\langle d_1(d_2 + d_2^c) \rangle $ provides the same result
as $ \langle {\cal G}_0
\rangle$. 
Thus, we may construct the corresponding current
\begin{equation}
{\cal G}_{{\rm inst},\mu}^\gamma({\bf p}_1,{\bf p}_2;P,Q)=i \langle d_1(p_1)
d_2(p_2+Q) j^{(2)}_\mu d_2(p_2) + d_1(p_1) d_2^{\tilde{c}}(p_2+Q)
j^{(2)}_{c,\mu} d_2^c(p_2) \rangle + (1 \leftrightarrow 2).
\end{equation}
When contracted with $Q^\mu$ this gives
\begin{eqnarray}
Q^\mu {\cal G}_{{\rm inst},\mu}^\gamma(p_1,p_2;Q)&=& i q_2[\langle d_1(p_1)
d_2(p_2) \rangle - \langle d_1(p_1) d_2(p_2 + Q) \rangle\nonumber\\
&+& \langle d_1(p_1) d_2^{c}(p_2) \rangle 
- \langle d_1(p_1) d_2^{\tilde c}(p_2+Q) \rangle] + (1 \leftrightarrow 2)\\
&=&q_2[\langle {\cal G}_0 \rangle({\bf p}_1,{\bf p}_2;P^0) 
- \langle {\cal G}_0 \rangle({\bf p}_1, {\bf p}_2 + {\bf Q};P^0 + Q^0)]
\nonumber\\
&& \qquad \qquad + (1 \leftrightarrow 2).
\end{eqnarray}
Consequently if a vertex function $\Gamma_{\rm inst}$ is constructed to be
a solution to Eq.~(\ref{eq:ET}) then the current matrix element defined by
\begin{equation}
{\cal A}_{{\rm inst},\mu}=\int \frac{d^3 p}{(2 \pi)^3} 
\bar{\Gamma}_{\rm inst}({\bf p} - {\bf Q}/2;P') 
{\cal G}_{{\rm inst}, \mu}^{\gamma (2)}({\bf p},{\bf P};Q)
\Gamma_{\rm inst}({\bf p};P) + (1 \leftrightarrow 2),
\label{eq:instAmu}
\end{equation}
obeys the current conservation condition (\ref{eq:instWTI}).  The
current ${\cal G}^\gamma_{{\rm inst},\mu}$ is simpler than the full ET
current and omits only effects stemming from retardation in the
current. Since our present calculations are designed to provide an
assessment of the role of negative-energy states and retardation
effects in the vertex functions, and these effects stemming from
retardation in the current are expected to be minor,
${\cal G}^\gamma_{{\rm inst},\mu}$ is used in these calculations.

\section{Electron-deuteron scattering: impulse approximation calculations}

\label{sec-results}

Now we are in a position to calculate the deuteron electromagnetic form 
factors $A$ and $B$, and the tensor polarization $T_{20}$. These are related
to the charge, quadrupole, and magnetic form factors of the deuteron,
$F_C$, $F_Q$, and $F_M$, by the following formulae,
\begin{eqnarray}
A&=&F_C^2 + \frac{8}{9} \eta^2 F_Q^2 + \frac{2}{3} \eta F_M^2,\\
B&=&\frac{4}{3} \eta (1 + \eta) F_M^2,\\
t_{20}&=& -\sqrt{2} \frac{x(x+2)+ y/2}{1 + 2(x^2 + y)};
\end{eqnarray}
where 
\begin{eqnarray}
x&=&\frac{2 \eta F_Q}{3 F_C},\\
y&=&\frac{2 \eta}{3} \left[\frac{1}{2} + (1 + \eta)
\tan^2\left(\frac{\theta_e}{2}\right)\right] \left(\frac{F_M}{F_C}\right)^2,\\
\eta&=&-\frac{Q^2}{4 M_D^2};
\end{eqnarray}
where $Q^2$ is the square of the four-momentum transfer. For elastic
electron-deuteron scattering, $Q$ is space-like and $\eta$ is positive.
In this work all plots and data are quoted at an electron angle of
$\theta_e=70$ degrees.

Deuteron form factors $F_C$, $F_Q$, and $F_M$ are related to the
matrix elements of the current ${\cal A}_\mu$ discussed in the previous
section, taken between the three different magnetic
quantum number states $|M \rangle=|+1 \rangle$, $|0 \rangle$, and $|-1
\rangle$ of the deuteron as follows:
\begin{eqnarray}
F_C&=&\frac{1}{3\sqrt{1 + \eta}e} (\langle 0|{\cal A}^0| 0 \rangle 
+ 2 \langle +1|{\cal A}^0|+1 \rangle),\\
F_Q&=&\frac{1}{2 \eta \sqrt{1 + \eta} e} (\langle 0|{\cal A}^0| 0 \rangle
- \langle +1|{\cal A}^0|+1 \rangle),\\
F_M&=& \frac{-1}{\sqrt{2 \eta (1 + \eta)}e} \langle +1|{\cal A}_+|0\rangle.
\end{eqnarray}

\begin{figure}[h]
\centerline{\BoxedEPSF{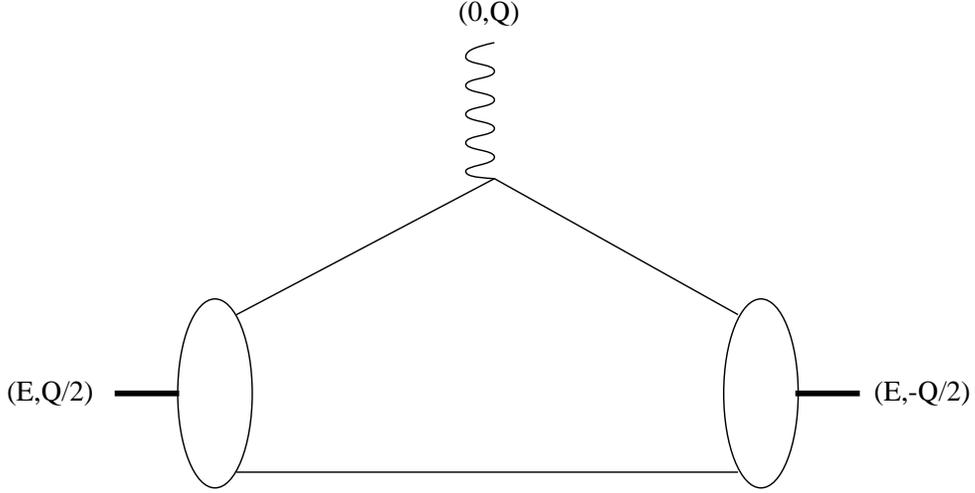 scaled 600}}
\vskip 5 mm
\caption{The kinematics of electron-deuteron scattering, for a momentum 
transfer $Q=(0,{\bf Q})$, in the Breit frame.}
\label{fig-Breit}  
\end{figure}

Matrix elements of ${\cal A}_\mu$ are calculated in the Breit frame,
with kinematics as shown in Fig.~\ref{fig-Breit}. In order to calculate
matrix elements such as those defined by Eq.~(\ref{eq:instAmu}) we require 
the vertex functions, or equivalently the wave functions, in the frame
where the total four-momentum is $P=(\sqrt{M_D^2 + {\bf Q}^2/4},{\bf Q}/2)$.
However, these are precisely the wave functions calculated in Section
\ref{sec-potentials}. Thus, we now take the wave functions 
constructed for the five different interactions of
Section~\ref{sec-potentials} at ${\bf P}^2$ ranging from 0 to 25 ${\rm
fm}^{-2}$ and insert them into the expression (\ref{eq:instAmu}). The
explicit form of the three-dimensional current ${\cal G}^\gamma _{{\rm
inst},\mu}$ is presented in Appendix~\ref{ap-iacurrentform}. In using
any of the interactions obtained with only positive-energy state
propagation we of course drop all pieces of the operator ${\cal
G}^\gamma_{{\rm inst},\mu}$ in negative-energy state sectors.

The general form of a matrix element of the operator ${\cal A}_\mu$,
using the deuteron vertex-function decomposition developed in
Ref.~\cite{WD95}, is, for coupling only to particle one,
\begin{eqnarray} 
\langle M' | {\cal A}_\mu | M \rangle=
\int \frac{d^3p}{(2 \pi)^3} [\Gamma_1^{\rho_1' s_1' \rho_2 s_2,M'} 
({\bf p} + {\bf Q}/2,P+Q)]^* \langle {\cal G}_0 \rangle^{\rho_1'
\rho_2}({\bf p} + {\bf Q}/2;P+Q)
\nonumber\\
J^{\rho_1' s_1' \rho_1 s_1 \rho_2 s_2}_{1,\mu} \langle {\cal G}_0
\rangle^{\rho_1 \rho_2}({\bf p};P) 
\Gamma_1^{\rho_1 s_1 \rho_2 s_2,M} ({\bf p};P)
+ (1 \leftrightarrow 2) ,
\label{eq:Jcompform}
\end{eqnarray}
with:
\begin{equation}
\langle {\cal G}_0 \rangle^{\rho_1 \rho_2} ({\bf p};P)
\equiv \rho_1 \rho_2 \frac{1}{\frac{1}{2}(\rho_1 + \rho_2)
P^0 - \epsilon_1 - \epsilon_2 + {1 \over 2} (\rho_1 - \rho_2)(\epsilon_1
- \epsilon_2) },
\end{equation}
and 
\begin{equation}
J^{\rho_1' s_1' \rho_1 s_1 \rho_2 s_2}_{1,\mu}=\rho_2
\bar{u}^{\rho_1' s_1'}(\rho_1' {\bf p}_1') 
\hat{J}_{1,\mu}^{\rho_1' \rho_1 \rho_2} u^{\rho_1 s_1}(\rho_1 {\bf p}_1),
\end{equation}
where the form of $\hat{J}_{1,\mu}^{\rho_1' \rho_1 \rho_2}$ is rho-spin
dependent and is given in Appendix~\ref{ap-iacurrentform}.

The single-nucleon currents used in these calculations are 
the usual one for extended nucleons, defined for $n=1$ or 2 by
\begin{equation}
j^{(n),\mu}=q_n\left(\gamma_n^\mu F^{(n)}_1(Q^2) +
\frac{i}{2M_N}\sigma_n^{\mu \nu} Q_\nu F^{(n)}_2(Q^2)\right).
\label{eq:extendednucleonj}
\end{equation}
The appearance of the factor $F_1$ in the first term here implies that
the WTI $Q^\mu j_\mu=\not \!\! Q$ is not satisfied. On the other hand,
as pointed out by Gross and Riska a current which obeys the original
WTI is
\begin{equation}
j_{WT}^\mu= \gamma^\mu + \frac{1}{2 M_N}\sigma^{\mu \nu} Q_\nu F_2(Q^2) + 
\left[F_1(Q^2) - 1 \right]\left[\gamma^\mu
- Q^\mu \frac{\gamma^\nu Q_\nu}{Q^2}\right]  .
\end{equation}
The difference between $j^\mu$ and $j_{WT}^\mu$ is proportional to
$Q^\mu$. Therefore, for elastic electron-deuteron scattering in the
Breit frame only $j^3$ will be affected if we adopt the WTI-obeying
current $j_{WT}^\mu$ instead of the original extended-nucleon current
$j^\mu$. Recall that in the Breit frame $j^3$ is formally constrained
by current conservation to be proportional to $j^0$, and is never
evaluated. Hence, it transpires that by formally adding a piece to the
one-body current $j^\mu$ we can obtain a one-body current for an
extended object which still satisfies $Q^\mu j_\mu=\not \!\! Q$, and
yet, in this calculation, leads to the same numerical results as
Eq.~(\ref{eq:extendednucleonj}). We choose to parametrize the
single-nucleon form factors $F_1$ and $F_2$ via the 1976 H\"ohler
fits~\cite{Ho76}.  Choosing different single-nucleon form factors does
not affect our qualitative conclusions, although it has some impact on
our quantitative results for $A$, $B$, and $T_{20}$.

Using such a one-body current the calculation described above
conserves the electromagnetic current if the vertex function
$\Gamma_1$ is obtained using an instant potential.  However, in all
other circumstances it violates the Ward-Takahashi identities by
omission of pieces that are required because of the inclusion of
retardation effects in the calculation. This violation of the WTI's
comes from two sources.

Firstly, in, for instance, the TOPT calculation, the pieces of the
current coming from terms in Eq.~(\ref{eq:A2mu}) of the form
$$
\langle G_0(P') (K^{(2)}-K_1^{(2)}) G_0(P) J_\mu G_0(P) \rangle
$$
are not included if Eq.~(\ref{eq:instAmu}) is applied. The inclusion
of these ``in-flight'' contributions to the current in our
calculations, and the consequent restoration of current conservation in
the TOPT calculation, will be the subject of a future paper. There
these contributions, and meson-exchange current effects such as the
$\rho \pi \gamma$ and $\omega \sigma \gamma$ MECs will be
calculated. Of course, similar ``in-flight'' pieces are missing from
the ``retarded ET'' calculation. 

Secondly, even if these ``in-flight'' pieces are included in the
retarded ET calculation the WTI's will not hold. Pieces will still be
missing from the current (\ref{eq:instAmu}).  Specifically, we need to
use the more complicated form of ${\cal G}^\gamma _{0,\mu}$ given in
Eq.~(\ref{eq:giprop}). However, calculations where $\langle
G_{C,\mu}^\gamma\rangle $ is omitted from the instant current matrix element
suggest that the total contribution of $\langle G_{C,\mu}^\gamma \rangle$ to
the observables is small.  Since $\langle G_{C,\mu}^\gamma\rangle$
itself makes a small contribution to observables, using a form of it
that only approximately satisfies current conservation is expected to
have an even smaller effect on our numerical results.

\section{Conclusion}

\label{sec-conclusion}

The ET formalism developed here provides a systematic
three-dimensional theory of electromagnetic interactions involving
relativistic bound states. This is achieved by integrating over time
components of relative momenta.  For the propagator of the theory,
this produces a form corresponding to zero relative time of the two
particles. In order for this formalism to incorporate the Z-graphs
that are expected in a quantum field theory, it is necessary for the
propagator to include terms that come from crossed Feynman graphs.
The predominant terms that are needed have been derived using a form
of the eikonal approximation. This leads to the ET propagator.  In
this paper, we discuss a refined version of the ET propagator
involving a different choice of $\kappa^0_2$ from that of
Refs.~\cite{MW87,WM89}.  This new choice of $\kappa_2^0$ avoids the
unphysical singularities which otherwise occur in the
three-dimensional ET interaction when it is evaluated in a frame where
the bound state has large three-momentum.

Given a suitable choice for the ET propagator, one may calculate the
interaction and the electromagnetic currents that must be used with
it. A full accounting of both the couplings to negative-energy states
and the role of retardations in the interaction is thereby obtained.

When the electromagnetic currents are constructed in our 3D theory the
necessary Ward-Takahashi identities are clearly satisfied provided
that one calculates both the the interaction and the electromagnetic
current to all orders. However, this is neither practical, nor, if an
effective hadronic Lagrangian is being used, desirable.  A
truncation of both the interaction and the electromagnetic current is
needed. It is not {\it a priori} evident that one may truncate the
expansion for the both the interaction and the electromagnetic current
in a way that systematically maintains current conservation.  We
show that this can in fact be done in the ET formalism.

Violations of Lorentz invariance occur when the interaction is
truncated.  For the range of momenta that arise in electron deuteron
scattering at TJNAF $Q^2$ these violations may be compensated for by
the use of a renormalized interaction.  It is found that the required
renormalization typically is no more than a few percent.  Based upon
this, we do not expect this violation of Lorentz invariance to have a
significant effect on results for electron-deuteron scattering.

Calculations have been performed for the impulse approximation, in
which we use an instant approximation for the electromagnetic current.
This current satisfies current conservation when used with deuteron
vertex functions that are obtained with instant interactions.  We also
have used this simple current with vertex functions that include the
full retardations obtained in the ET formalism.

Impulse approximation results fall systematically below
experimental data for the form factors $A$ and $B$ for $Q >$ 3--4
fm$^{-1}$.  This deficiency of the theoretical calculations at higher
$Q$ indicates that additional mechanisms beyond the impulse
approximation should be significant.  However, the existing tensor
polarization data are reasonably well described, and this is
consistent with previous analyses that have shown $T_{20}$ to be less
sensitive to two-body currents.

The role of negative-energy states is found to be not very large.  Our
impulse-approximation numerical results are in closer agreement with
those of Hummel and Tjon~\cite{HT89,HT90,HT94} than with those based
upon the spectator formalism~\cite{vO95}.  Because the ET formalism
incorporates the relevant Z-graphs in a preferable way, we are
confident that Z-graphs play a minor role in calculations that are
based upon standard boson-exchange models of the $NN$ interaction.
The role of retardation corrections in the deuteron vertex functions
also is rather minor.

Further calculations are needed in order to incorporate the full ET
current and the meson-exchange currents.  

\acknowledgements{We thank the U.~S. Department of Energy
for its support under grant no. DE-FG02-93ER-40762. D.~R.~P.
is grateful for the warm hospitality of the Special Research
Centre for the Subatomic Structure of Matter, where the 
writing of this paper was completed.}

\appendix 

\section{Leading effects in $1/M$ of time-ordered Z-graphs}

\label{ap-ZgraphTOPT}

Here we examine the fourth-order graphs for the $NN$ t-matrix in 
the equal-time formalism and show that the contributions of order 
$1/M$ are correctly reproduced in the ET equation with instant interactions.

\begin{figure}[h]
\centerline{\BoxedEPSF{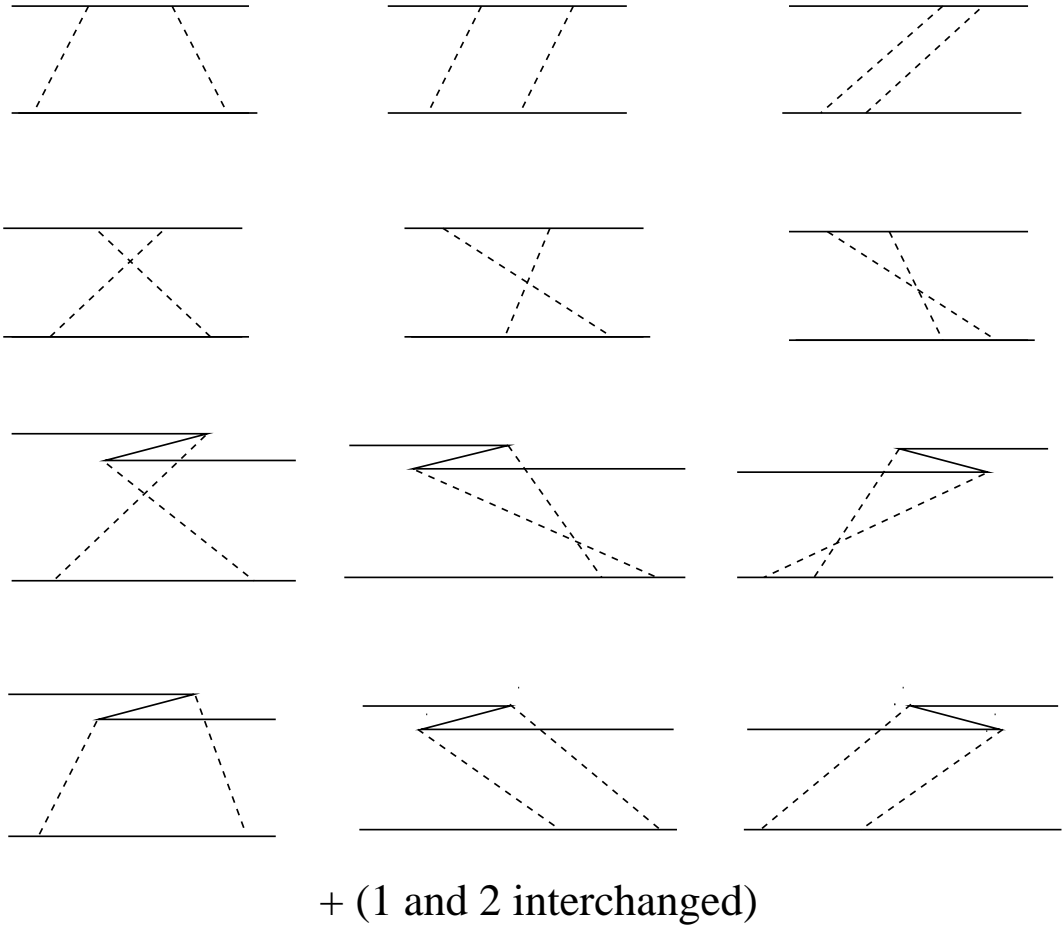 scaled 700}}
\vskip 5 mm
\caption{The graphs which contribute to the three-dimensional $NN$
t-matrix at fourth order in the coupling, up to order $1/M$. The
graphs on the first and third line come from the $NN$ Feynman box
graph and those on the second and fourth line come from the $NN$
Feynman crossed-box graph.}
\label{fig-ZgraphTOPT}  
\end{figure}

The fourth-order piece of the equal-time Green's function is:
\begin{equation}
\langle G_0 K^{(2)} G_0 K^{(2)} G_0 \rangle + \langle G_0 K^{(4)} G_0 \rangle
\label{eq:fourthorder}
\end{equation}
where $K$ is the full Bethe-Salpeter kernel. If we consider only
positive-energy states on the external lines, and decompose the
internal nucleon lines according to the different rho-spins which are
possible, we produce forty-eight graphs. However, if we restrict our
attention to graphs which contribute at order $1/M$ or above only the
twelve graphs shown in Fig.~\ref{fig-ZgraphTOPT} are relevant. Of
these, those on the first and third lines arise from the first,
iterative, part of expression (\ref{eq:fourthorder}) while those on
the second and fourth line come from the crossed-box piece of
Eq.~(\ref{eq:fourthorder}). 

The external Green's functions are now amputated, and the static
limit of
\begin{equation}
T_{++ \rightarrow ++}^{(4)}=\langle G_0 \rangle_{++}^{-1}
\left[\langle G_0 K^{(2)} G_0 K^{(2)} G_0 \rangle + \langle G_0 K^{(4)} G_0 
\rangle\right]_{++ \rightarrow ++} \langle G_0 \rangle_{++}^{-1}
\end{equation}
is taken.  In this limit the first two graphs in
Fig.~\ref{fig-ZgraphTOPT} are infinitely enhanced, as they come
from the iteration of the lowest-order interaction $K_1^{(2)}$.
Meanwhile the next four graphs give energy denominators:
\begin{equation}
- \frac{1}{\omega_1} \frac{1}{\omega_1 + \omega_2} \frac{1}{\omega_2}
- \frac{1}{\omega_2} \frac{1}{\omega_1 + \omega_2} \frac{1}{\omega_1}
- 2 \frac{1}{\omega_1}\frac{1}{\omega_1 + \omega_2}\frac{1}{\omega_1},
\end{equation}
where $\omega_1$ ($\omega_2$) is the energy of the first (second)
meson emitted from nucleon one.  These graphs do not vanish in the
limit $M \rightarrow \infty$. They correspond to a shorter-range
interaction than $K_1^{(2)}$, and are formally part of the
fourth-order three-dimensional kernel $K_1^{(4)}$. Their effects are
usually subsumed into $K_1^{(2)}$ through the fitting of the
parameters in that interaction to the $NN$ scattering data.

The contribution of the remaining six graphs begins at order $1/M$. At that
order they give energy denominators
\begin{eqnarray}
&-& \frac{1}{\omega_1}\frac{1}{2 M} \frac{1}{\omega_2} 
- \frac{1}{2 M} \frac{1}{\omega_1 + \omega_2}  \frac{1}{\omega_2}
- \frac{1}{\omega_1} \frac{1}{\omega_1 + \omega_2} \frac{1}{2 M} \nonumber\\
&& \qquad \qquad - \frac{1}{\omega_2} \frac{1}{2 M} \frac{1}{\omega_1}
- \frac{1}{2 M} \frac{1}{\omega_1 + \omega_2} \frac{1}{\omega_1}
- \frac{1}{\omega_2} \frac{1}{\omega_1 + \omega_2} \frac{1}{2 M}.
\end{eqnarray}
Including the appropriate spinors and factors to obtain the full
expression shows that the sum of these graphs yields:
\begin{equation}
g^4 \bar{u}_1^+ V_1(1) u_1^- \bar{u}_2^+ V_1(2) u_2^+
\bar{u}_1^- V_2(1) u_1^+ \bar{u}_2^+ V_2(2) u_2^+
\left(-\frac{1}{\omega_1^2}\right)\left(-\frac{1}{2 M}\right)
\left(-\frac{1}{\omega_2^2}\right),
\label{eq:1/Mbits}
\end{equation}
where $V_i(j)$ is the vertex for the interaction of meson $i$ with
nucleon $j$. Note that in writing this expression we have taken some
liberties with the ordering of these vertices. However, the error that
this produces is of higher order in $1/M$. To order $1/M$
Eq.~(\ref{eq:1/Mbits}) reproduces the $++ \rightarrow ++$ matrix element
of the iteration of the instant interaction using the $-+$ sector of the ET
propagator:
\begin{equation}
\bar{u}_1^+ \bar{u}_2^+ K^{(2)}_{\rm inst} \frac{\Lambda_1^- \Lambda_2^+}{-2M}
K^{(2)}_{\rm inst} u_1^+ u_2^+.
\label{eq:iteratively}
\end{equation}
Similarly, the $1 \leftrightarrow 2$ version of expression
(\ref{eq:iteratively}) gives the six fourth-order particle-two
Z-graphs correctly to order $1/M$.

Thus we see that using the ET propagator with an instant interaction
correctly reproduces the $1/M$ pieces of the fourth-order Z-graphs in
the scattering series. By contrast, the Gross or BSLT propagator with
an instant interaction would yield a fourth-order Z-graph contribution
which is only half the correct size. 

\section{Explicit form of the impulse approximation current}

\label{ap-iacurrentform}

In this appendix we present the explicit form of the impulse approximation
current which yields a current-conserving calculation if vertices 
obtained using an instant interaction are considered. Recall that
\begin{equation}
{\cal G}_{{\rm inst},\mu}^\gamma({\bf p}_1,{\bf p}_2;P,Q)=i \langle
d_1(p_1+Q) j^{(1)}_\mu d_1(p_1) d_2(p_2) + d_1^{\tilde{c}}(p_1+Q)
j^{(1)}_{c,\mu} d_1^c(p_1) d_2(p_2) \rangle + (1 \leftrightarrow 2).
\end{equation}
Inserting the forms of the propagators and performing the
integrals over relative momenta we arrive at a form which
is most easily expressed in terms of the quantity $\hat{J}_\mu^{\rho_1' \rho_1
\rho_2}$, as written in Eq.~(\ref{eq:Jcompform}). 
In rho-spin conserving sectors it is:
\begin{equation}
\hat{J}^{\rho_1 \rho_1 \rho_2}_\mu=j_\mu -
(1 - \delta_{\rho_1 \rho_2}) \tilde{j}_\mu,
\end{equation}
where $\tilde{j}_\mu$ is the object defined for nucleon two in
Eq.~(\ref{eq:tildej2}).  In our calculation the $\tilde{j}_\mu$
defined there is multiplied by $F_1(Q^2)$ in order to take into
account the extended-nucleon structure. Likewise, $j_\mu$ is the usual
extended-nucleon current (\ref{eq:extendednucleonj}).

Meanwhile, in rho-spin changing sectors we have:
\begin{eqnarray}
{\hat J}^{-++}_\mu&=&\left[1 - 2\frac{\kappa_2^{0 \prime} - \epsilon_2}{Q^0 +
\epsilon_1 + \epsilon_1'} +
\frac{P^0 - \epsilon_1 - \epsilon_2}{Q^0 + \epsilon_1 + \epsilon_1'} 
+ \frac{P^0 - \epsilon_1 - \epsilon_2}{Q^0 + 2({\kappa_1^0} - \kappa_1^{0 \prime}) - \epsilon_1 -
\epsilon_1'} \right] j_\mu\nonumber\\ &-& \frac{P^0 -
\epsilon_1 - \epsilon_2}{Q^0 + 2({\kappa_1^0} -
\kappa_1^{0 \prime}) - \epsilon_1 - \epsilon_1'} \tilde{j}_\mu\\
{\hat J}^{+-+}_\mu&=&\left[1 + 2\frac{\kappa_2^0 - \epsilon_2}{Q^0 -
\epsilon_1 -
\epsilon_1'} - 
\frac{{P^0}' - \epsilon_1' - \epsilon_2}{Q^0 + 
2(\kappa_1^0 - {\kappa_1^0}') + \epsilon_1 + {\epsilon_1}'} -
\frac{{P^0}' - \epsilon_1' - \epsilon_2}{Q^0 - \epsilon_1 - \epsilon_1'} \right] j_\mu \nonumber\\
&+& \frac{P^{0 \prime} - \epsilon_1' - \epsilon_2}{Q^0 + 2(\kappa_1^0 -
{\kappa_1^0}') + \epsilon_1 + \epsilon_1'} \tilde{j}_\mu\\
{\hat J}^{-+-}_\mu&=&\left[1 + 2\frac{\kappa_2^0 + \epsilon_2}{Q^0 + \epsilon_1 +
\epsilon_1'} - \frac{P^{0 \prime} + \epsilon_1' + \epsilon_2}{Q^0 +
2(\kappa_1^0 - {\kappa_1^0}') - \epsilon_1 - \epsilon_1'} 
- \frac{P^{0 \prime} + \epsilon_1' + \epsilon_2}{Q^0 + \epsilon_1 + \epsilon_1'}\right] j_\mu \nonumber\\
&+& \frac{P^{0 \prime} + \epsilon_1' + \epsilon_2}{Q^0 + 2(\kappa_1^0 -
\kappa_1^{0 \prime}) - \epsilon_1 - \epsilon_1'} \tilde{j}_\mu\\
{\hat J}_\mu^{+--}&=&\left[1 - 2\frac{\kappa_2^{0 \prime} +
\epsilon_2}{Q^0 - \epsilon_1 - \epsilon_1'} + 
\frac{P^0 + \epsilon_1 + \epsilon_2}{Q^0 -
\epsilon_1 - \epsilon_1'} + \frac{P^0 + \epsilon_1 + \epsilon_2}{Q^0 + 2({\kappa_1^0} -
\kappa_1^{0 \prime}) + \epsilon_1 + \epsilon_1'}\right]j_\mu \nonumber\\
&-& \frac{P^0 + \epsilon_1 + \epsilon_2}{Q^0 + 2({\kappa_1^0} -
\kappa_1^{0 \prime}) + \epsilon_1 + \epsilon_1'} \tilde{j}_\mu
\end{eqnarray}
These reduce to the formulae of Devine~\cite{WD94} if the appropriate
limits are taken. Note that in $++$ states the current $\hat{J}_\mu$
is just the usual single-nucleon current (\ref{eq:extendednucleonj}).

\newpage
\begin{table}[h]
\caption{Meson quantum numbers, masses, cutoffs, and couplings as taken
from the Bonn-B model. Note that the number in brackets in the $\rho$ row is
the $\rho$-meson tensor coupling.}
\label{table-mesondata}
\begin{tabular}{|c|c|c|c|c|c|}
\tableline
Meson   & $J^P$ & $T$ & Mass (MeV) & Cutoff (MeV) & $g^2/4 \pi$\\ \tableline 
$\pi$   & $0^-$ &  1  &   138.03   &   1200       &     14.6   \\ \tableline
$\eta$  & $1^-$ &  0  &   548.8    &   1500       &      5.0   \\ \tableline
$\rho$  & $1^+$ &  1  &   769.0    &   1300       &   0.95 (6.1)\\ \tableline
$\omega$& $1^+$ &  0  &   782.6    &   1500       &     20.0   \\ \tableline
$\delta$& $0^+$ &  1  &   983.0    &   1500       &    3.1155  \\ \tableline
$\sigma$& $0^+$ &  0  &   550.0    &   2000       &            \\ \tableline
\end{tabular}
\end{table}
 
\begin{table}[h]
\caption{Sigma coupling required to produce the correct deuteron binding
energy in the five different models under consideration here.}
\label{table-sigmacoupling}
\begin{center}
\begin{tabular}{|c|c|c|}
\tableline
Interaction  &  States included & $g_\sigma^2/4 \pi$ \\ \tableline
Instant      &       ++         &      8.08          \\ \tableline
Klein        &       ++         &      9.64          \\ \tableline
Retarded ET  &       ++         &      8.39          \\ \tableline
Instant ET   &      All         &      8.55          \\ \tableline
Retarded ET  &      All         &      8.44 	     \\ \tableline
\end{tabular}
\end{center}
\end{table}
\newpage

\newpage

\end{document}